\documentclass[useAMS,usenatbib,usegraphicx]{mn2e}

\def \kms {km s$^{-1}$}

\title[New observations of the
Seyfert galaxy Mrk 315]{New photometric and spectroscopic observations
of the Seyfert galaxy Mrk 315}
\author[S. Ciroi et al.]{S. Ciroi,$^{1,3}$\thanks{E-mail: ciroi@pd.astro.it}
V.L. Afanasiev,$^{2}$
A.V. Moiseev,$^{2}$
V. Botte,$^{1,3}$
F. Di Mille,$^{1}$
\newauthor
S.N. Dodonov,$^{2}$
P. Rafanelli,$^{1}$
A.A. Smirnova$^{2}$\\
$^{1}$Department of Astronomy, Padova University, Vicolo dell'Osservatorio 2,
Padova, 35122, Italy\\
$^{2}$Special Astrophysical Observatory, Nizhnij Arkhyz, 369167 Russia\\
$^{3}$ Guest investigator of the UK Astronomy Data Centre}

\begin{document}


\maketitle

\label{firstpage}

\begin{abstract}
We present new important results about the intermediate-type Seyfert galaxy
Mrk 315, recently observed through optical imaging and integral-field
spectroscopy. Broad-band images were used to study the morphology of the host
galaxy, narrow-band H$\alpha$ images to trace the star forming regions, and
middle-band [O\,{\sc iii}] images to evidence the distribution of the highly
ionised gas. Some extended emission regions were isolated and their physical
properties studied by means of flux calibrated spectra. High resolution
spectroscopy was used to separate different kinematic components in the
velocity fields of gas and stars. Some peculiar features characterise this
apparently undisturbed and moderately isolated active galaxy. Such features,
already investigated by other authors, are re-analysed and discussed in the
light of these new observations. The most relevant results we obtained are:
the multi-tiers structure of the disc; the presence of a quasi-ring of
regions with star formation much higher than previous claims; a secondary
nucleus confirmed by a stellar component kinematically decoupled by the main
galaxy; a new hypothesis about the controversial nature of the long filament,
initially described as hook-shaped, and more likely made of two independent
filaments caused by interaction events between the main galaxy and two dwarf
companions.
\end{abstract}


\section{Introduction}

Understanding the processes responsible for triggering the activity in
galactic nuclei is one of the fundamental outstanding questions regarding
Active Galactic Nuclei (AGNs). Several mechanisms have been invoked during the
last decades, as the presence of circumnuclear star clusters,
nuclear bars, discs or spirals, and the interaction between galaxies, in form
of close encounters and mergers. Unfortunately, none of them produced conclusive
results. For example, nuclear bars and spirals seems not to be more common among
AGNs than among non-active galaxies \citep{car98, pg02, mart03}.
Numerical simulations showed that the gravitational interaction between galaxies
can bring gas from the disc toward the nuclear regions \citep{mh94, hm95, mh96}.
Nevertheless, statistical studies of the large-scale environments of nearby
AGNs produced until now controversial results
\citep[see e.g.][]{da84, bush86,fws88, mac89, lauri94, raf95, derob98, sch01},
and failed to demonstrate that
a one--to--one relationship between activity and interaction does really exist.
On the contrary deep high resolution imaging and
follow--up spectroscopy of these galaxies allowed to identify merging
systems or the presence of close faint companions, suggesting
that the investigation of the interaction--activity relation should be
addressed to AGN hosts and their immediate surroundings.
Therefore, one of the most natural approaches toward this topic is looking for
effects of interaction in apparently isolated and morphologically undisturbed
nearby Seyfert galaxies.
In this paper we present new results about the intermediate-type Seyfert
galaxy, Mrk 315.

Mrk 315 (II Zw 187) is a well known active galaxy.
Spectroscopically classified as Seyfert 1.5 by \citet{kos78}, it was studied
by \citet{wil88} with high resolution long-slit scanning of the nuclear and
extra-nuclear regions of the galaxy (within a radius of 5 arcsec). His spectra,
obtained around H$\beta$ and [O\,{\sc iii}] emission lines, revealed the
presence of a two distinct kinematic components associated to gas showing
different ionisation degrees.\\
Mrk 315 was also extensively investigated by \citet{mac86}, \citet{mac94} and
\citet{sm01}. These authors found an unusual jet-like ionised gas feature,
emitting in [O\,{\sc iii}] narrow-band filter but not visible in VLA
radio maps, and suggested for it to be gas drawn out from Mrk 315 by a tidal
interaction with another galaxy.
By means of HST images they discovered a knot close to the active nucleus,
which they interpreted as a remnant nucleus, proposing a scenario of `...{\it
an AGN caught in the act of the initiation stage of a tidally induced
feeding}'. \\
Finally, Mrk 315 was observed in optical, infrared and radio wavelength
domains by \citet{non98}. They confirmed the presence of a secondary
nucleus embedded in a chain of faint structures surrounding the active nucleus,
and discussed the possible origin of these structures in favour of the
\citet{mac94} merger hypothesis.

Here we present new photometric and spectroscopic data of Mrk 315,
which confirm partly the previous findings mentioned above and show for
the first time that we are observing the effects of the interaction of two
dwarf galaxies with Mrk 315.
The paper is organised as follows: the photometric and spectroscopic data
are presented in Section 2; the morphology of Mrk 315 is investigated in
Section 3; the 2D-spectrophotometric data are analysed in Section 4
with the aim to study in detail the physical properties of gas in the
circum-nuclear regions of the galaxy; the stellar and gaseous kinematics is
presented and discussed in Section 5; in Section 6 we analyse the environment
surrounding Mrk 315 to look for objects physically connected to the
galaxy; finally, in Section 7 we discuss the results obtained for each of the
features identified in Mrk 315.

\section[]{Observations and Data Reduction}

\subsection[]{Photometry}

\begin{figure*}
\centerline{\includegraphics[width=10 cm]{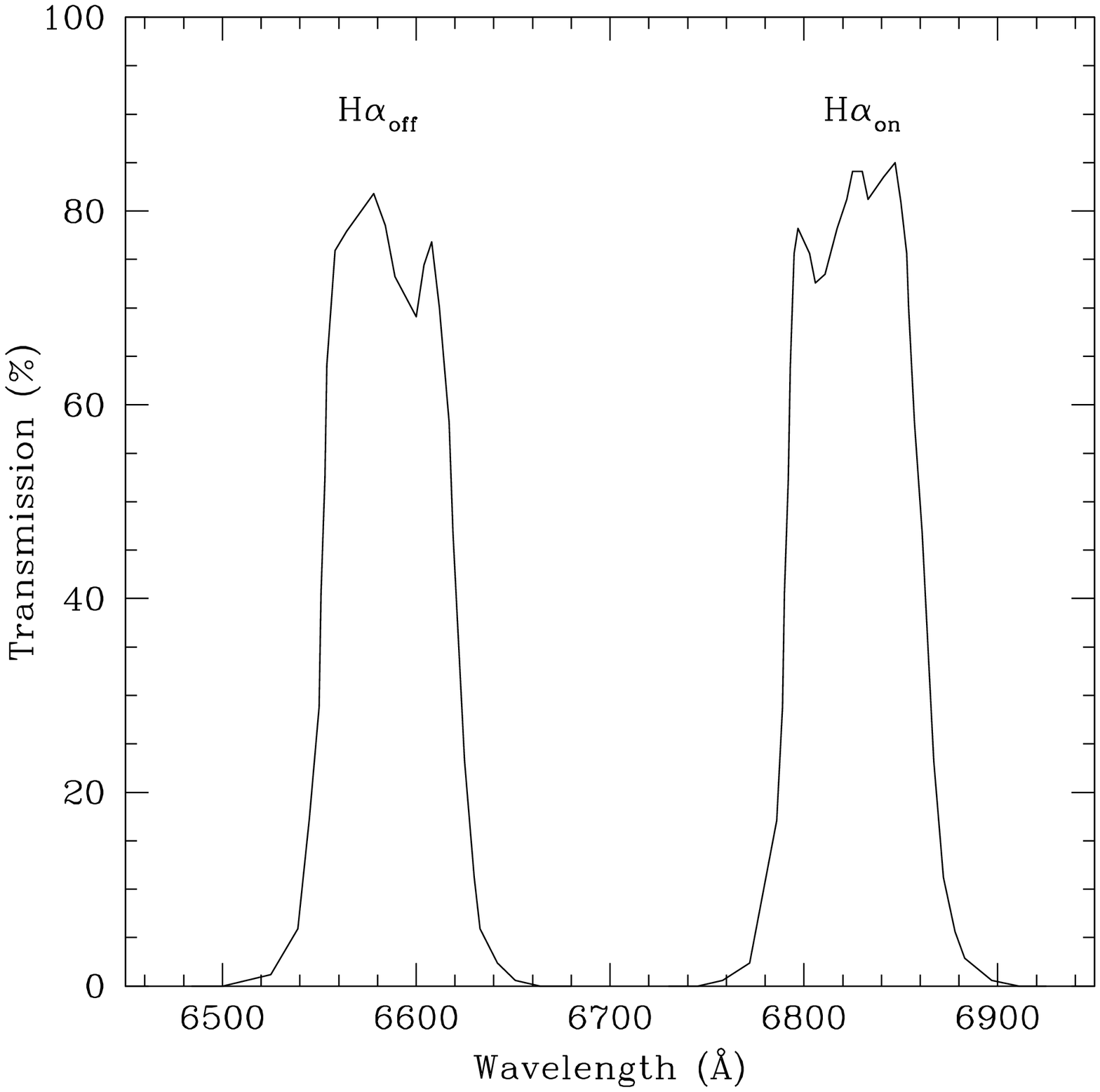}}
\caption{Transmission curves of the VATT narrow-band filters H$\alpha_{on}$ 
and H$\alpha_{off}$ with central wavelength $\lambda_c$=6830 \AA, and 
$\lambda_c$=6580 \AA\ respectively. 
Both filters have similar width $\Delta\lambda\sim70$ \AA .} 
\label{vattfilt}
\end{figure*}

Broad-band B and R, and narrow-band H$\alpha$ ($\lambda_c$=6830 \AA,
$\Delta\lambda\sim70$ \AA; H$\alpha_{on}$ hereafter), H$\alpha$-continuum
($\lambda_c$=6580 \AA, $\Delta\lambda\sim70$ \AA; H$\alpha_{off}$ hereafter)
images were obtained at the 1.8-m Vatican Advanced Technology Telescope (VATT)
of the Vatican Observatory (Arizona, USA) in October 1998, with a 2k$\times$2k
CCD camera. The transmission curves of the narrow-band filters are plotted
in Fig.~\ref{vattfilt}.
Middle-band [O\,{\sc iii}] (filter SED520, $\lambda_c$=5229 \AA,
$\Delta\lambda=310$ \AA; [O\,{\sc iii}]$_{on}$ hereafter),
[O\,{\sc iii}]-continuum (filter SED574, $\lambda_c$=5730 \AA,
$\Delta\lambda=160$ \AA; [O\,{\sc iii}]$_{off}$ hereafter), broad-band V and
R$_c$ images were obtained in August 2002 and September 2004 with SCORPIO
\citep{am05}, a multi-mode focal reducer mounted at the 6-m telescope (BTA) of
the Special Astrophysical Observatory (SAO-RAS, Russia).
See Table \ref{imalog} for observation details.
These images were reduced with IRAF\footnote{IRAF is distributed by
the National Optical Astronomy Observatories, which are operated by
the Association of Universities for Research in Astronomy, Inc.,
under cooperative agreement with the National Science Foundation}
(for VATT images) and with IDL (for BTA image) in a standard way by
subtracting bias and correcting for flat-field variations and cosmic
ray events.
Also B, V, R$_c$ and I$_c$ images from \citet{chatzi00}, available in
NED (NASA Extragalactic Database), were involved in our study.
The calibration in the magnitude scale for filters R$_{c}$
and I$_{c}$ was obtained by means of a bright star, located 30 arcsec
South-East of the galaxy, whose photometry was carried out by \citet{bsd00}.
Instead, the calibration for filters B and V was carried out according to
published aperture photometry.
The six best aperture measurements out of the eight
available in database HyperLeda\footnote{http://leda.univ-lyon1.fr/} have been
selected. Finally, for the image obtained with filter [O\,{\sc iii}]$_{off}$
the calibration was carried out by using the V band.
The accuracy of zero-point in any case is better than 0.1 mag.

\begin{table*}
\caption{Photometric observations.}
\label{imalog}
\begin{tabular}{@{}llllllll@{}}
\hline
Date & T$_{exp}$ & Filter & seeing & Scale & FOV & Instr. & Tel. \\
& (sec) & & (arcsec) & (arcsec/px) & (arcmin) & & \\
\hline
1998 Oct. 29 &  1200 & B          & 1.5 & 0.4 & 6$\times$6 & CCD Camera & VATT \\
             &   600 & B          & &               & &     &      \\
             &  1200 & R          & &               & &     &      \\
             &   600 & R          & &               & &     &      \\
             &  1200 & H$\alpha_{on}$ & &            & &     &      \\
             &  1200 & H$\alpha_{off}$ & &           & &     &      \\
2002 Aug. 30 &  1800 & [O\,{\sc iii}]$_{on}$  & 1.6 & 0.29 & 5$\times$5 & SCORPIO & BTA \\
             &  1800 & [O\,{\sc iii}]$_{off}$ & &     & &     &     \\
2004 Sep. 08 &  1320 & V          & 1.6 & 0.35 & 6$\times$6 & SCORPIO & BTA \\
             &   600 & R$_c$      & 1.5  &       & & &  \\
\hline
\end{tabular}
\end{table*}

\subsection{Panoramic spectroscopy at the 6m telescope}

\subsubsection[]{Integral-Field Spectrograph MPFS}

Mrk 315 was observed in December 2000 and August 2003 with the MultiPupil
Fiber Spectrograph (MPFS), the integral-field unit mounted at the primary
focus of the 6-m telescope \citep{adm01}.
The MPFS takes simultaneous spectra from 240 spatial elements (constructed in
the shape of square lenses) that form on the sky an array of $16\times15$
elements (in 2000), or from 256 spatial elements ($16\times16$ in 2003).
The angular size is 1 arcsec/element. The detector
was a TK1024 in 2000 and EEV CCD42-40 ($2048 \times 2048$ px) in 2003.
A description of the MPFS is available at SAO RAS web page:
{\it http://www.sao.ru/hq/lsfvo/devices.html}.

During the first run, the galaxy was observed at low resolution
for spectrophotometric purposes, while the second run was addressed to the gas
and stellar kinematics investigation, and the galaxy was observed at higher
resolution.
The log of MPFS observations is given in Table \ref{speclog}.

The data were reduced by using the software developed at the
SAO RAS by V.L. Afanasiev and A.V. Moiseev and running in the IDL
environment. The primary reduction included bias subtraction,
flat-fielding, cosmic-ray hits removal, extraction of individual
spectra from the CCD frames, and their wavelength calibration using
a spectrum of a He-Ne-Ar lamp. Subsequently, we subtracted the
night-sky spectrum from those of the galaxy. The spectra of
spectrophotometric standard stars were used to convert counts
into absolute fluxes.

\begin{table*}
\caption{Spectroscopic observations.}
\label{speclog}
\begin{tabular}{@{}llllllll@{}}
\hline
Date & T$_{exp}$ & Sp. Range & Sp. Res. & Disp. & Seeing &  Instr. &  Mode \\
& (sec) & (\AA) & (\AA) & (\AA/px) &  (arcsec) & &  \\
\hline
1994 Aug. 17 &  1500     &    4700 - 5700   &  2.2  & 0.8  & 1.2  &  ISIS    &  \\
2000 Dec. 03 &  3600     &    3700 - 6100   &  7.5  & 2.6  & 1.6  &  MPFS    & \\
             &  4800     &    5000 - 7400   &  7.5  & 2.6  & 1.6  &  MPFS    & \\
2003 Aug. 24 & 14400     &    4860 - 6330   &  4.2  & 0.75 & 1.5  &  MPFS    & \\
2002 Nov. 02 &  30 x 250 & [O\,{\sc iii}]    &  1.2  & 0.54 & 2.0  &  SCORPIO & IFP \\
2002 Nov. 07 &   900     &    3700 - 7300   &  10.0 & 1.8  & 1.5  &  SCORPIO & slit\\
\hline
\end{tabular}
\end{table*}

\subsubsection{Scanning Fabry-Perot Interferometer}

Mrk 315 was observed in November 2003 with SCORPIO \citep{am05} in Fabry-Perot
mode. The Queensgate interferometer Fabry-Perot (IFP) ET-50 was used at
the 300th interference order (for the redshifted [O\,{\sc iii}] $\lambda4959$
line). The free spectral range between neighbouring orders (interfringe) was
about 16 \AA.
The detector used was an EEV CCD42-40 operated
with $2\times2$ binning for reading-out time saving. The IFP
was installed into the parallel beam, inside the focal reducer
SCORPIO which provides a field-of-view of $\sim 6$ arcmin and a spatial
scale of $\sim 0.3$ arcsec/px. A brief description of this device is also given
on the SAO RAS web page: {\it http://www.sao.ru/hq/moisav/scorpio/scorpio.html}.
In addition, the IFP mode in SCORPIO is described by \citet{moi02}.
The log of the IFP observations is presented in Table \ref{speclog}.

To reduce the interferometric observations we used a custom development
software \citep{moi02}, running in the IDL environment. After the primary
reduction (bias, flat-field, cosmic hits),
we removed the night-sky spectrum, converted
the data to the wavelength scale, and prepared them as `data cubes'.
The data cube was smoothed by a Gaussian function with FWHM = 2 px in the
spectral and spatial domain under the ADHOC \footnote {ADHOC software was
written by J. Boulesteix (Observatoire de Marseille).
See http://www-obs.cnrs.fr/ADHOC/adhoc.html} package.
The spatial resolution after smoothing was about 2.5 arcsec.
The velocity fields of the ionised gas, and images in the emission line
[O\,{\sc iii}] $\lambda 4959$ were mapped by means of a Gaussian fitting of
the emission line profiles. Moreover, we created the images of the galaxy in
the `red' continuum close to the emission line.

\subsubsection{Long-slit spectroscopy}

A long-slit spectrum obtained with ISIS Double Beam Spectrograph at the William
Herschel Telescope on August 1994 was extracted from the ING public archive.
The grating R600B was used in combination with a 1.5 arcsec-slit, at position
angle PA= 114\degr\ .

We also obtained a long-slit spectrum of a source located $\sim 1$ arcmin
South-East of Mrk 315 with SCORPIO in November 2002.
The detector used was the EEV CCD42-40 2048$\times$2048 pixels
and the slit had a width of 1.2 arcsec.

Data reduction was performed using our IDL-based software for SCORPIO data, and
IRAF for ISIS data. We followed the standard procedures applied to long-slit
spectra: bias and flat-field correction, two-dimensional wavelength calibration,
flux calibration, included atmospheric extinction correction, and night-sky
background subtraction.

\section{Morphology}

\subsection{Isophotal analysis}

\begin{figure*}
\centerline{\includegraphics[width=12 cm]{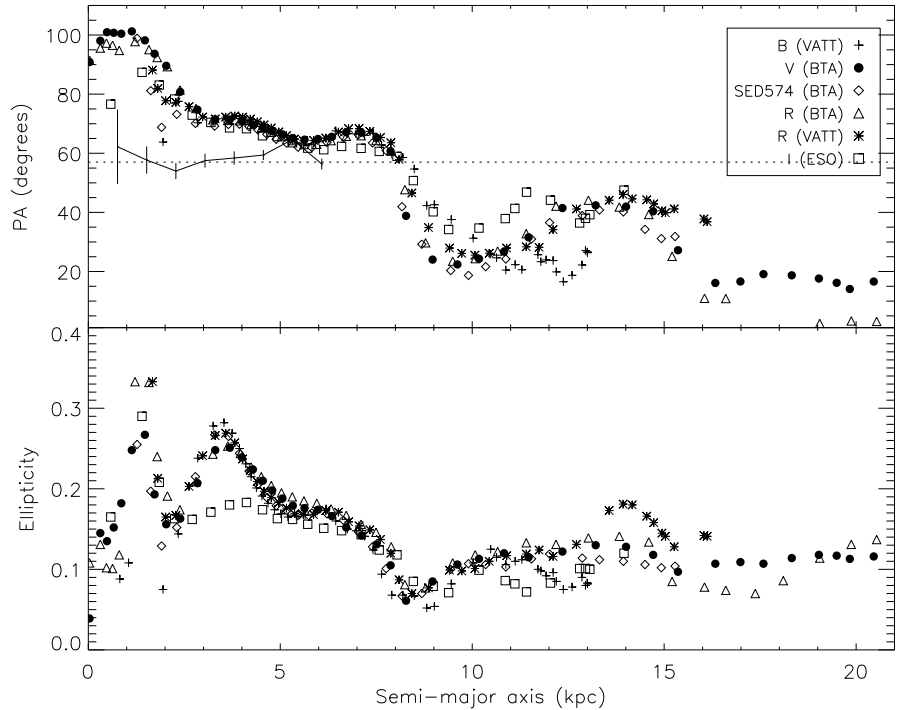}}
\caption{Elliptical isophote parameters from the available
images of Mrk 315. In the upper panel the position angle is plotted
vs. the semi-major axis in kpc. The symbols correspond to different optical
bands. The solid line marks the dynamical axis measured on the velocity
field of stars (Fig.~\ref{starvel}), the dashed line corresponds to the mean 
value of the dynamical axis (which is accepted as line of nodes). 
The ellipticity of the isophotes is plotted in the lower panel.} 
\label{ell_param}
\end{figure*}

The morphology of the galaxy was investigated by fitting parametric ellipses
to its isophotes in all of the images at our disposal, and
fixing the centre at the location of the nucleus.
Then, the plots of ellipticity ({\it e}) and position angle (PA) were
obtained as a function of the semi-major axis length, expressed in
kpc (see Fig.~\ref{ell_param}).
The spatial scale  is 0.75 kpc/arcsec assuming a
distance to the galaxy of 153.6 Mpc, directly measured on our spectroscopic
data (V$_{sys}=11517\pm9$, see Section 5), and H$_0 = 75$ \kms\ Mpc$^{-1}$.
These plots show a first peak in ellipticity between 1 and 2 kpc and a second
peak between 3 and 4 kpc.  Then, the isophotes have a smooth trend up to
the outer regions with ellipticity ranging around $e \sim 0.1$.
A slight dip in ellipticity between 8 and 9 kpc corresponds to a
change in isophote orientation from PA$\sim60^\circ-70^\circ$
(between 3 and 8 kpc) to PA$=15^\circ-40^\circ$ (between 10 and 20 kpc).

\subsection{2-D decomposition}

Mrk~315  is classified in NED (NASA Extragalactic Database) as an elliptical
galaxy (type E1), and so the distribution of the surface brightness in its
external regions should be described  by a spheroidal component.
On the other hand, \citet{chatzi01} claimed for an extended external disc in
Mrk~315, and stressed that the photometric parameters of the galaxy
do not match its morphological type.
We tried to decompose the surface brightness distribution by applying
`standard' components: an exponential disc with central brightness $\mu_d$ and
scale length $r_d$, and a S\'ersic bulge \citep{ser68} with effective radius
$r_{eff}$,  effective brightness $\mu_{eff}$ and power index $n$ varying from 1
to 4. In addition, a point-like source profile was included to account for the
presence of the AGN.

The 1-D surface brightness profile and the radial behaviour of
the isophotes are too complex to be decomposed in a
unique way. A more stable and reliable result could be obtained with the
method of the 2-D decomposition, which uses the whole bi-dimensional
information of the surface brightness distribution \citep[see e.g. ][]{mgh98}.
We applied the IDL-based software GIDRA, developed at SAO RAS by A.V.
Moiseev, and we followed the same iterative method of consecutive comparisons
between 1-D mean profiles and 2-D models, which allowed \citet{moi04} to study
effectively the morphology of double-barred galaxy candidates
\citep[see also ][]{sa00,sva01}.
The seeing-convolved decomposition was carried out for images in all
available bands, and among them we chose deeper BTA data in the
filters [O\,{\sc iii}]$_{off}$, V and $R_c$, because  the images from
\citet{chatzi00} reach a lower limit of magnitude, and the
VATT images show a significant gradient of the sky background around the
galaxy.
The limit surface brightness of the BTA broad-band images, corresponding
to signal-to-noise ratio S/N=3, is $\mu_V=26.0$ mag arcsec$^{-2}$ and
$\mu_{R_c}=25.7$ mag arcsec$^{-2}$.

Firstly, we obtained the azimuthal average of the outer isophotes, whose
position angle and ellipticity are expected to correspond to the disc
orientation, and we calculated the approximated values of $\mu_d$ and $r_d$ by
fitting the brightness profile in this disc-dominated region. Then,
the 2-D model of this exponential disc was subtracted from the
original image. The residual image was averaged in annular apertures
and the mean profile was fitted by a Se\'rsic-law. Finally, the
central peak of brightness was fitted by a two-dimensional
Gaussian function, with FWHM similar to the PSF of the image, which corresponds
to the active nucleus. The 2-D model bulge+core was subtracted from the
original image and the cycle was repeated to refine the model.

The model subtracted images are shown in Fig.~\ref{mod}.
In both cases we have detected two characteristic features on these images.
First of all, a faint two armed spiral structure is visible in the outer part
of the galaxy in the interval $r=7-18''$ ($\sim 5-14$ kpc).
Second, the residual image has a significant excess of brightness at $r<8-10''$,
which originates a `bump' in the mean profile.
The brightness excess can be equally explained with two different orientations
of the outer disc, but in this case the kinematic properties of the stars must
be involved in our analysis (see Section 7.1).
We tried to fit the brightness distribution in this region with an additional
inner exponential disc.
The resulting parameters of the decomposition are listed in
Table~\ref{tab_decomp}. The contribution of the components to the
total brightness profile are shown in Fig.~\ref{mod}.

\begin{table*}
\caption{Parameters of the 2D decomposition of the brightness
distribution} \label{tab_decomp}
\begin{tabular}{cllllllll}
\hline
Band          & \multicolumn{3}{c}{bulge}& \multicolumn{2}{c}{outer disc}& \multicolumn{2}{c}{inner disc} & Lum. ratio\\
              &$\mu_{eff}$   &$r_e$    & n&$\mu_{d1}$&$r_{d1}$ &$\mu_{d2}$&$r_{d2}$& $L_B/L_D$\\
              & (mag)        & (kpc)   &  & (mag)    & (kpc)   &   (mag)  & (kpc)   &  \\
 \hline
[O\,{\sc iii}]$_{off}$&20.82$\pm0.33$&2.01$\pm0.14$&2.0$\pm0.3$&22.35$\pm0.04$&6.99$\pm0.07$&20.06$\pm0.22$&2.05$\pm0.08$&0.62$\pm0.13$\\
    V        &21.70$\pm0.18$&2.55$\pm0.13$&3.0$\pm0.3$&22.79$\pm0.05$&7.07$\pm0.10$&19.58$\pm0.10$&1.90$\pm0.04$&0.56$\pm0.12$\\
 R$_c$        &21.09$\pm0.21$&2.49$\pm0.15$&3.0$\pm0.3$&21.85$\pm0.05$&6.16$\pm0.08$&19.03$\pm0.06$&1.98$\pm0.03$&0.52$\pm0.13$\\
     \\
    V         &19.69$\pm0.07$&3.05$\pm0.30$&1.7$\pm0.4$&23.10$\pm0.05$&7.60$\pm0.11$&     &    &2.67$\pm0.35$ \\
 R$_c$         &19.75$\pm0.10$&2.35$\pm0.04$&1.5$\pm0.2$&21.77$\pm0.04$&6.31$\pm0.05$&     &    &2.47$\pm0.29$\\
 \hline
\end{tabular}
\label{tab_phot2}
\end{table*}

\subsection{Peculiar features}

The brightness excess corresponds to the complex structure surrounding
the nucleus, already discovered by \citet{mac94} and \citet{non98}, and having a
quasi-ring shape.

Comparing the H$\alpha$ continuum-subtracted image with the
model subtracted V image (V$_{res}$), we observe that the
emission is extended like this quasi-ring and, even if smoothed, it follows
very well its layout, which is therefore characterised by star formation
(Fig.~\ref{Ha+O3}a).
Embedded in this structure is a bright knot (named knot K throughout this paper)
marked with `X' on Fig.~\ref{Ha+O3}.
This knot is the brightest feature, after the active nucleus, observed
on the V$_{res}$ image, but it does not appear as an
equally bright H$\alpha$ source.
Conversely, it is well evident in the I-band HST image, published by
\citet{mac94}, proving its prevailing stellar nature.
In fact this knot is the supposed
secondary nucleus of the merger hypothesis claimed by these authors.

The [O\,{\sc iii}] continuum-subtracted image  shows a strongly different
distribution of the higher ionisation gas, which seems only weakly to follow
the quasi-ring (Fig.~\ref{Ha+O3}b). Of course this is not unexpected, since
[O\,{\sc iii}] emission is usually weak in regions, where thermal ionisation by
star formation dominates.
A prominent feature radially extended from the nucleus up to $\sim 7$ kpc
does not correspond to any other structure, suggesting the
idea of gas directly ionised by the active nucleus.
Moreover, it is interesting to point out that the putative secondary nucleus
has no evident, or maybe very weak, counterpart in [O\,{\sc iii}]
emission.

Further, low-brightness structures at S/N$\sim3-5$ are
visible in the deep V and R$_c$ images obtained at BTA (Fig.~\ref{imaV}).
We observe a sort of trail apparently non co-planar to the disc of the galaxy
and wounded around it, and a plume located South-West. At least two stellar
shells can be identified $\sim 15$ kpc North and $\sim 30$ kpc North-West of
the galactic disc. Such shells are similar to those discovered in the merger
galaxy Mrk 298 \citep{rad04}.
In addition, two filaments are laid in the surroundings of Mrk 315.
The first one (F1 hereafter), already identified by \citet{mac86}, extends up to
$\sim 70$ kpc from the central regions of the galaxy at PA $\sim 145\degr$.
The second one (F2 hereafter) seems to begin $\sim 75$ kpc North-West of
the galaxy. Then, it crosses F1 in projection, extends at PA
$\sim 120\degr$ toward the North-East side of the galaxy, where it turns to
PA $\sim 0\degr$, and finally ends at the location of an extended source
$\sim 45$ kpc South-East of Mrk 315.
The total projected extension of F2 is $\sim 140$ kpc.

\section[]{Spectrophotometric Data Analysis}

The low resolution spectra were analysed with IRAF tasks.
A correction for Galactic absorption, A(V)= 0.68, was applied and the
atmospheric absorption band at 6860-6920 \AA\ was removed by means of the flux
calibrated spectrum of the standard star.

Several emission lines are present in Mrk 315 spectra.
To perform a more precise measurement of their fluxes, these lines were
corrected for the underlying stellar contribution. An IRAF task made by us on
the basis of the insights given by \citet{hfs93} allows to subtract the spectrum
of a template early-type galaxy after having conveniently rescaled its continuum
and diluted its absorption features \citep[see][for a more extensive discussion
about this step]{cir03}.

A first reconstruction of the galaxy image in the field-of-view of the
spectrograph was done convolving each spectrum with the transmission curve of a
typical Johnson V filter. This image together with narrow-band images obtained
integrating in a small wavelength range around H$\alpha$ and [O\,{\sc iii}]
$\lambda$5007 allowed to locate the position of the nucleus in the
brightest pixel. This position does not coincide with that shown by a similarly
reconstructed image in [O\,{\sc ii}] $\lambda$3727, because of the
atmospheric refraction effect. To solve the problem we calculated the
correction function to be applied to each image before any comparison.
This was done simply producing several narrow-band continuum images covering
both spectral ranges and measuring the centroid of the galaxy
in each image with a bi-dimensional Gaussian fitting.
Finally the centroid positions were interpolated separately as a
function of wavelengths.

Since Mrk 315 is an intermediate-type  Seyfert galaxy, its nuclear spectrum
shows permitted lines with composite profiles: a narrow component emitted by the
Narrow-Line Region (NLR) and a broad component emitted by the Broad-Line Region
(BLR). Given its sub-parsec size the BLR can be considered as a point-like
source, therefore its emission will not be limited to the nuclear spectrum, but
distributed in the surrounding regions depending on the seeing
during the observations.
In order to obtain spectra with only NLR emission lines, we applied the
following procedure.
Firstly, a hand-made PSF was reproduced in the field-of-view of MPFS with a
bi-dimensional normalised Gaussian function centred on the brightest pixel
(in the H$\alpha$ reconstructed map), where the BLR should be located, and with
a FWHM similar to the seeing value.
Then, the spectrum of the brightest pixel was measured.
A deblending of the H$\alpha$+[N\,{\sc ii}] $\lambda\lambda6548,6583$ profile
by means of an iterative multi-Gaussian fitting procedure involving
four components, H$\alpha$-broad (H$\alpha_b$), H$\alpha$-narrow (H$\alpha_n$),
[N\,{\sc ii}] $\lambda6548$ and [N\,{\sc ii}] $\lambda6583$, allowed us to
build a spectrum showing only H$\alpha_b$. This spectrum was
replicated to cover the entire field-of-view, and the 240 H$\alpha_b$
lines were rescaled to obtain the spatial distribution of the BLR
emission in H$\alpha$. This was done by multiplying each of them for the
scaling factor given by the pixel intensity of the PSF at the
corresponding location on the field-of-view.
Finally these spectra were subtracted from those containing both BLR and NLR
emission lines.\\
Two parameters were fine tuned to better remove the broad component: the
FWHM of the PSF and its position within the pixel where the active nucleus is
located. Obviously a bi-dimensional Gaussian is only an approximation of the real
PSF and moreover the low spatial resolution (1 arcsec) implies that even an
off-centring of 0.2 px can distribute the flux in a significantly different
way. These are the reasons why we could not fix these two parameters a-priori.
The same steps were applied to H$\beta$.

After having removed the broad emission lines, the spectra of Mrk 315 were
measured. An IRAF task made by us and based on a recursive application of
NGAUSSFIT allowed to interactively fit each emission line with
Gaussian functions, and obtain tables with fundamental parameters, as position, flux,
FWHM, amplitude and EW.
These tables can be quickly used to reconstruct maps of the 16$\times$15
resolution elements for whatever of these parameters.
Relative errors were also calculated by estimating the
RMS of the continuum close to each emission line. In most of the spectra, the
fluxes of the brightest lines have errors below 30 per cent (Table \ref{flerr}).

\begin{table*}
\caption{Flux errors of the measured emission lines.}
\label{flerr}
\begin{tabular}{@{}cccccccccc@{}}
\hline
[O\,{\sc ii}] 3727 &  [O\,{\sc iii}] 4363 & He\,{\sc ii} 4686 & H$\beta$ &
[O\,{\sc iii}] 5007 & [O\,{\sc i}] 6300 & H$\alpha$ & [N\,{\sc ii}] 6583 &
[S\,{\sc ii}] 6724 & $\Delta$I/I \\
\hline
42 & 5  & 0  & 65 & 61 & 10 & 73 & 65 & 57 & $<$0.10 \\
80 & 24 & 12 & 82 & 82 & 53 & 88 & 83 & 73 & $<$0.20 \\
96 & 38 & 53 & 94 & 89 & 74 & 94 & 92 & 83 & $<$0.30 \\
\hline
\end{tabular}

\medskip
Columns (1)--(6) show the percentage of spectra whose relative flux errors
are below the values given in column (7).
\end{table*}

\subsection{Emission line ratios}

We have reconstructed maps of the most important emission lines visible in the
spectra of Mrk 315, i.e. [O\,{\sc ii}]$\lambda$3727, [O\,{\sc
iii}]$\lambda$4363, He\,{\sc ii}$\lambda$4686, H$\beta$, [O\,{\sc iii}]$\lambda$5007,
[O\,{\sc i}]$\lambda6300$, H$\alpha$, [N\,{\sc ii}]$\lambda6583$
and [S\,{\sc ii}]$\lambda6716+31$ (also named [S\,{\sc ii}]$\lambda6724$
throughout this paper).
These maps, grouped in
Fig.~\ref{allmaps}, show the spatial distribution of the ionised gas.
Maps of H$\beta$ and H$\alpha$ lines have a smoothed shape, where structures are
not easy to be distinguished because of the low spatial resolution.
Nevertheless, a significant improvement is reached by overlapping the contours
of the V$_{res}$ image (Fig.~\ref{regHa+O3}a).
These contours helped us to identify integral-field spectra belonging to
specific regions. In particular we defined the nucleus (N),
three regions of the quasi-ring structure (A1, A2, A3), numbered
counter-clockwise from South-West to North-East, and finally the knot (K)
located 2.5 arcsec East of the nucleus. \\
A similar procedure was used for the [O\,{\sc iii}] map, whose shape is
well in agreement with the contours of the [O\,{\sc iii}]
continuum-subtracted image (Fig.~\ref{regHa+O3}b). On this map we have isolated other two
regions, one radially extended from the nucleus and North-West oriented (J1),
the other smaller and fainter, located $\sim 6$ arcsec East of the nucleus (J2).
Finally, [O\,{\sc i}] and [S\,{\sc ii}] maps clearly show a
secondary bright emission region identified with the knot K.\\
In Fig.~\ref{1d} we plotted the integrated spectra of these identified regions.

Then we have calculated the emission line ratio maps on the basis of the
diagnostic diagrams of \citet[][VO hereafter]{vo87}, i.e. [O\,{\sc iii}]/H$\beta$,
[N\,{\sc ii}]/H$\alpha$, [O\,{\sc i}]/H$\alpha$, and
[S\,{\sc ii}]/H$\alpha$. These ratios are useful to study the ionisation
of gas and identify the kind of ionising source. As done before, each ratio map
was compared with a contour ([O\,{\sc iii}] continuum-subtracted image,
or V$_{res}$ image) to make the analysis clearer (Fig.~\ref{ratios}).
[N\,{\sc ii}]/H$\alpha$ and [O\,{\sc i}]/H$\alpha$ ratios show
peaked values at the location of the knot K. This is visible also in the
[S\,{\sc ii}]/H$\alpha$ map, where a larger structure extends
clockwise from K toward West in the outer part of A3. High values of
[S\,{\sc ii}]/H$\alpha$ ratio are also located South of A1.
The [O\,{\sc iii}]/H$\beta$ map shows two loci where the gas is highly
ionised, except for the nucleus: J2 and the outer regions of J1.\\
We have also reconstructed the map of the [O\,{\sc ii}]/[O\,{\sc iii}] emission
line ratio (Fig.~\ref{o2o3}), which
is abundance independent and a good indicator of the ionisation degree.
Unlike the pairs of emission lines used by the VO diagnostic diagrams,
which are close in wavelength so that their ratios are almost unaffected by
dust extinction, the [O\,{\sc ii}]/[O\,{\sc iii}] ratio must
be reddening corrected. Therefore we have firstly calculated the observed
H$\alpha$/H$\beta$ ratios and applied a theoretical Balmer decrement of 2.86
to all emission regions but the nucleus N, where a 3.1 was used following the
suggestions by \citet{ost89}. As extinction law we used that given by
\citet{ccm89}.
The map of the visual absorption A$\rm _V$  (Fig.~\ref{de}, right) shows
clearly that the major extinction is distributed in a region
including N, K and J2 with values ranging from 2 to 3 mag (light grey).
Lower values (A$_V \sim$ 1--1.5) are typical of A1, A2 and A3, while in J1 the
extinction is negligible.\\
The map of the reddening corrected [O\,{\sc ii}]/[O\,{\sc iii}] ratios
shows two zones (dark grey) located on the nucleus N and on J1, where the
[O\,{\sc iii}] is dominant ([O\,{\sc ii}]/[O\,{\sc iii}]
$<$ 1) and the ionisation is mainly produced by the AGN.
The [O\,{\sc ii}] is stronger than [O\,{\sc iii}]
([O\,{\sc ii}]/[O\,{\sc iii}] $>$ 3--10) roughly in correspondence of the
other regions A1, A2, A3, and K (light grey), thus
indicating the prevailing thermal origin of the ionisation sources.\\
Finally, the map of the electron density ($n_e$) distribution,
given by the [S\,{\sc ii}]6716/6731 ratio (Fig.~\ref{de}, left), shows
the highest values ($\sim 10^3$ cm$^{-3}$) where also the
extinction is higher, while it assumes values in the range
100--300 cm$^{-3}$ North of the nucleus, including A3, and even less than
100 cm$^{-3}$ in A1 and A2.

To carry out a more detailed analysis we plotted the emission line ratios
[O\,{\sc iii}]/H$\beta$ versus [N\,{\sc ii}]/H$\alpha$,
[O\,{\sc i}]/H$\alpha$, and [S\,{\sc ii}]/H$\alpha$ respectively,
(Figs.~\ref{vo}a, \ref{vo}b, \ref{vo}c) according to the VO diagrams.
We added the empirical
borderlines, which separate AGN from H\,{\sc ii} regions (solid line),
and AGN from LINER or supernova remnants (SNR; dashed line), or in other words
define the ranges of ratios for which non-thermal, thermal or shock ionisation
dominates. An inspection of the diagrams suggests
that the observed ratios are consistent with the structure identified in the
image analysis (see Section 3). In particular spectra belonging to N (open circles)
fall in the AGN zone, although three of them are close to the AGN/LINER
borderline. Interestingly, all spectra around the central one (the brightest
pixel in e.g. H$\beta$, H$\alpha$ or [N\,{\sc ii}] maps of Fig.~\ref{allmaps})
where the active nucleus is expected to be located, show higher
[N\,{\sc ii}], [O\,{\sc i}] and [S\,{\sc ii}] to
H$\alpha$ values.
On the contrary, spectra of K (filled circles) fall in the LINER/SNR
zone and show a lower ionisation degree ([O\,{\sc iii}] $\approx$
H$\beta$). Thanks to the relatively good seeing, N and K spectral properties
are well discriminated even if they are close to each other.
The easiest and maybe
more corrected interpretation is that the gas in region K is dominated by shock
effects. Nevertheless, we cannot rule out the possibility that K itself is an
active secondary nucleus, namely a LINER.
A1 and A3 have most of their line ratios typical of H\,{\sc ii} regions
(open squares and crosses respectively),
while A1 has some close to the AGN/H\,{\sc ii} borderline and A3 shows
some values of [O\,{\sc i}] and [S\,{\sc ii}] to
H$\alpha$ falling in LINER/SNR zone.
These last points are located among regions N, K and A3 and their line ratios
strongly indicate that the gas is compressed and shock ionisation occurs.
Region A2 has line ratios mostly falling in the H\,{\sc ii} region zone
(filled squares), but two of them are close to the AGN/H\,{\sc ii} borderline.
These two spectra of A2 are just those near the nucleus N, and therefore are
probably `polluted' by AGN radiation. Indeed it was shown by \citet{rhr98} that
an increasing contribution to the AGN ionisation by star forming regions pushes
the emission line ratios from the AGN toward the H\,{\sc ii} zone.
J1 has five spectra in common with A2 and one in common with A1.
In agreement with what we said before about J1 and the [O\,{\sc iii}]/H$\beta$
map, the inner parts of J1 (skeletal triangles) are characterised by lower
ionisation, while the outer parts fall in the AGN zone.
This is not unexpected because the
overlap between J1 and A2 makes lower the emission line ratio
[O\,{\sc iii}]/H$\beta$ since H$\beta$ is strong in star forming regions.
The outer parts of J1 ($> 7$ arcsec) are reached by AGN radiation, and so J1
can be considered an ionisation cone. It is likely that J1 and A2 overlap only
in projection on the sky, but are not coplanar.
The other cone could be located in region J2
(open stars), whose spectra have also AGN-like properties.
Its small extension and relatively
faintness could be caused by the fact that we are observing it through the
disc of the galaxy.

The remaining emission regions have line ratios mostly filling the H\,{\sc ii}
zone in all VO diagrams, while few points have AGN properties, in particular
those around J1, therefore characterised by diffuse
AGN radiation, and finally others fall in the LINER/SNR zone.
A more careful identification of these last points, carried out by
reconstructing the map of their positions within the field of view, reveals
that they form a ring around the identified regions.
This ring is made mostly by regions whose spectra
have LINER/SNR properties only in the [S\,{\sc ii}]/H$\alpha$ diagram,
while few regions are present in two diagnostics or in all the three.
In addition two adjacent regions not connected to the ring have a
LINER/SNR identification in all the three plots, and are located among N, A1
and A2.
The shape and displacement of these regions suggests that shocks
caused by gas compression around the inner quasi-ring structure could be at the
origin of their emission line ratios. Nevertheless a single
plot based indication makes this hypothesis not completely convincing, since
the regions whose spectra show higher values of
[S\,{\sc ii}]/H$\alpha$ ratio could simply have an overabundance of
[S\,{\sc ii}], even if in this case the ring-like distribution of
these regions remains unexplained.

\subsection{Star Formation Rate and Energy Budget}

We have used the H$\alpha$ emission to estimate the star formation rate (SFR)
in the field of view of integral-field data.
The H$\alpha$ reddening corrected fluxes were first converted into
luminosities, then the total luminosities of the portion of galaxy observed
with MPFS, and of each region identified on the H$\alpha$ emission line map,
were calculated.
Finally, we applied the formula by
\citet{ken98}: $\rm SFR~(M_{\odot}~yr^{-1}) = 7.9\times10^{-42}~L_{H\alpha}$.
In Table~\ref{halpha} we list these values, together with the surface density
of SFR in $\rm M_{\odot}~yr^{-1}~pc^{-2}$ units.
The total $\rm L_{H\alpha}$ includes also the nuclear contribution, which is
clearly AGN dominated and therefore the SFR value is overestimated.
Removing $\rm L_{H\alpha}(N)$, which accounts for about 22 per cent of the
total $\rm L_{H\alpha}$, we obtain SFR $\rm \sim 35~M_{\odot}~yr^{-1}$ and
$\rm \Sigma_{SFR}\sim 4.36\times10^{-7}~M_{\odot}~yr^{-1}~pc^{-2}$.

For consistence with these results, we tried to estimate SFR by using other two
indicators, the infrared and radio emissions.
We used the IRAS data of Mrk 315 extracted from the Point Source Catalogue, and
in particular the 60 $\rm \mu m$ flux, S$_{60}=1.505$ Jy. Then, we followed the
suggestions given by \citet{ch00}, based on \citet{rr97}, who made use of the
60 $\rm \mu m$ luminosity to estimate the total Far Infrared emission within the range
1--1000 $\rm \mu m$, $\rm L_{FIR} \sim 1.7 \times L_{60}~erg~sec^{-1}$, and
converted into star formation rate with the formula:
$\rm SFR~(M_{\odot}~yr^{-1})=L_{FIR}/2.2 \times 10^9~L_{\odot}$.
We obtained the following values, $\rm L_{FIR} (1-1000~\mu m) =
3.32\times10^{44}~erg~sec^{-1}$ and $\rm SFR \sim 39~M_{\odot}~yr^{-1}$.
Also in this case, the total FIR emission includes the AGN contribution,
therefore this value of SFR should be considered as an upper limit.\\
The radio emission value at 1.425 GHz was taken from \citet{non98}, who measured
the total flux and the contribution from the nucleus and the knot separately.
We removed the nuclear emission obtaining S$_{20}=19$ mJy, and luminosity $\rm
L_{20}=5.38\times10^{22}~W~Hz^{-1}$.
This value was then converted into SFR using the formula given by
\citet{bell03}, and obtaining $\rm \sim 30~M_{\odot}~yr^{-1}$, in good agreement
with the SFRs found by using H$\alpha$ and FIR emission.

From the radio luminosity we also calculated the supernova rate (SN), since at
1.425 Ghz the radio emission is usually dominated by non-thermal radiation
(synchrotron) produced by electrons accelerated by supernova remnants and
explosions. To obtain the non-thermal luminosity $\rm L_{NT}$, we followed
the assumption by \citet{cy90} and \citet{bell03} that the thermal radio fraction 
at 1.425 GHz is about 10 per cent. Then, we applied the relation:
$\rm L_{NT} (W~Hz^{-1})\sim 1.3\times10^{23}~(\nu/1~GHz)^{-\alpha}~\nu_{SN}~
(yr^{-1})$, given by \citet{cy90}, and we adopted the spectral
index $\alpha=0.9$ for the overall galaxy \citep{non98}.
The SN rate ($\nu_{\rm SN}$) is about 0.5 yr$^{-1}$, a value much higher
than that expected in normal spiral galaxies, 1 SN/100 yrs per 10$^{10}$
$\rm L_{\odot}(B)$ \citep[see e.g. ][]{bt91}, but consistent with the rates
observed in starburst galaxies, which are typically in the range 0.1-1.0
yr$^{-1}$ \citep{sll98,man03,nut04}.
The equally high SFR is in agreement with observations of starburst galaxies,
since in the local Universe mildly obscured and UV luminous starbursts show
rates of 5-50 $\rm M_{\odot}~yr^{-1}$ \citep{heck05}.

Since the VO diagrams indicate that the J1 region, especially its outer part,
is highly ionised, it is interesting to evaluate whether the nuclear radiation
can sustain this observed high ionisation degree.
To carry out this test, we have firstly isolated a region of 3$\times$2 arcsec,
located $\sim 6$ arcsec NW of the nucleus, within J1 and showing a spectrum
dominated by AGN excitation according to the diagnostic
diagrams. In the approximation that the AGN is the only source producing the
H$\alpha$ photons observed in this region, we converted its total H$\alpha$
luminosity into the number of ionising photons
$\rm Q_{ion}=7.3\times10^{11}~L_{H\alpha}$ \citep{ost89}. Then we corrected this
number for the effect of the geometrical dilution of the nuclear radiation
flowing through the solid angle $\Omega$ subtended by the region,
$\rm Q_{nuc}=(4\pi/\Omega) \times Q_{ion}\sim3\times10^{54}~photons~sec^{-1}$.

Now we use this rough estimate of the number of ionising photons emitted by the
source, to calculate the expected ionisation parameter
$\rm U=Q_{nuc}/(4\pi r^2 N_{H} c)$
at the distance of the region. Assuming r $\sim$ 4.6 kpc and $\rm N_{H} \sim
10^2$ cm$^{-3}$ (obtained from the [S\,{\sc ii}]6716/6731 ratio), we obtain $\rm
log~U \sim -3.4$.
This value does not agree with observed line ratios. A simple comparison with
published photoionisation models \citep[see e.g. ][]{hsf93}, indicates that this
expected ionisation parameter is typical of a low ionisation degree.
Indeed, the [O\,{\sc ii}]/[O\,{\sc iii}] emission line ratio $\sim 0.4-0.5$
(reddening corrected), observed both in the nucleus and in the region should
correspond to a value of $\rm log~U \sim -2.5$ \citep{pen90,ks97}.
In the end, the nuclear ionising radiation alone seems not enough to produce the
high excitation levels detected in J1, especially in its outer parts.
Another source of high energy photons must be present, like hot stars, or more
likely shock waves.
A similar result was obtained by \citet{mac86}, who followed a different
approach.

\begin{table}
\caption{H$\alpha$ luminosities and star formation rates.}
\label{halpha}
\begin{tabular}{cccc}
\hline
 Region & $\rm L_{H\alpha}$ & SFR  & $\Sigma_{\rm SFR}$ \\
 & ($10^{42}$ erg sec$^{-1}$) & ($\rm M_{\odot}~yr^{-1}$) &
 ($\rm M_{\odot}~yr^{-1}~pc^{-2}$) \\
\hline
 N  & 1.30 & ... & ... \\
 K  & 0.50 & 3.9 & $1.13\times10^{-6}$ \\
 A1 & 1.14 & 9.0 & $5.79\times10^{-7}$ \\
 A2 & 0.47 & 3.7 & $8.10\times10^{-7}$ \\
 A3 & 0.84 & 6.6 & $8.19\times10^{-7}$ \\
 Total & 5.84 & 46 & $5.32\times10^{-7}$ \\
\hline
\end{tabular}

\medskip

\end{table}

\section[]{Kinematics}

The kinematics of Mrk 315 was investigated essentially by means of the
integral-field MPFS and Fabry-Perot data.

\subsection[]{Fabry-Perot data}

The  kinematics of the highly ionised gas was studied in detail through the
spectral analysis of the [O\,{\sc iii}] emission line.
The high resolution of the Fabry-Perot data allowed to clearly separate two
distinct components in the emission line profile: a first component belonging
to the disc of the galaxy and showing a pattern of almost circular rotation
with velocities ranging from 11350 and 11750 \kms, and a second `redder'
component corresponding to the previously mentioned J1 and J2 regions, and
characterised by high velocity gas radially moving toward the outskirts of the
galaxy. The difference to the systemic velocity is $\sim + 500$ \kms.
In addition, the first kinematic component shows a clear nuclear outflow with
V$\sim 11480$ \kms (Fig.~\ref{o3vel_fp}).\\
We remark that the disc circular rotation seen in
[O\,{\sc iii}] agrees with the circular rotation observed in $H\alpha$ and
in the velocity field of stars (see next section).

The analysis of the long-slit spectrum, even if at lower resolution,
showed the same two kinematic components both in
[O\,{\sc iii}] and also in H$\beta$.
We fitted separately these components, and then we compared the
velocity curves with the Fabry-Perot data extracted at the same PA of the slit,
obtaining a good agreement.

We compared also our results with those published by \citet{wil88}, who
found the two kinematic components in [O\,{\sc iii}], one
characterised by rotational motions, and the other extended from the nucleus
toward North and North-West with an unclear kinematical structure.
Unlike \citet{wil88}, we observe clearly the bleuward and redward components
both in [O\,{\sc iii}] and H$\beta$.
Moreover, we observe a larger splitting of the two kinematic components than
\citet{wil88}, maybe beacuse of an higher spectral resolution which allowed us
to obtain more precise measurements of the line profiles.

\subsection[]{MPFS data}

The low resolution MPFS spectra were used to derive the overall
kinematics of both low and high ionisation gas within the field-of-view.
Firstly the wavelength positions of H$\alpha$ emission lines were measured by
means of Gaussian fittings and then converted into heliocentric velocities.
Later we constructed the maps of the line-of-sight velocity fields of the
brightest emission lines. The absolute accuracy of the velocity measurements,
evaluated from the air-glow emission line wavelengths, was about 10-15 \kms.
A continuum map was also obtained by adding the fluxes in the
spectral ranges free from emission lines (5600-5900 \AA).

Velocity maps show a deviation from circular rotation in oxygen
([O\,{\sc i}], [O\,{\sc ii}], [O\,{\sc iii}]) and sulphur ([S\,{\sc ii}]) lines.
In particular, in [O\,{\sc ii}] we can see a sudden increase of velocity
with $\Delta V \sim +300$ \kms, where the knot K is located.
A sudden decrease of velocity ($\Delta V \sim -200$ \kms) in [O\,{\sc iii}]
confirms the outflow already shown by Fabry-Perot and long-slit data.

The kinematics of the stellar component in Mrk 315 was studied by means of
the high resolution MPFS data. The line-of-sight
velocity and dispersion velocity fields were constructed.
We used  the `classic' cross-correlation
method adapted for working with MPFS spectra \citep{moi02}. The region
5050-6050 \AA\, containing numerous stellar absorption
features (Mg\,{\sc i}, Fe\,{\sc i}, Na\,{\sc i} etc.) was analysed.
As a template for cross-correlation the spectra of the twilight sky, observed
in the same night as the galaxy, were used.
The estimated errors were
$\sim$ 10 \kms\ and 10-20 \kms\ for the velocity and velocity dispersion
respectively.
The line-of-sight velocity distribution (LOSVD) of the stars in the central
region of the galaxy shows two clearly peaked structures, with a separation of
about 600 \kms. By means of a double-Gaussian fit of the LOSVD we have
constructed maps of line-of-sight velocity and  velocity dispersion for
both components.  The results are shown in Fig.~\ref{starvel}.
Both components have velocity fields with a pattern corresponding to circular
rotation of inclined discs. The centre of rotation of the `blue' LOSVD component
coincides with the photometric nucleus of Mrk 315, and the centre of rotation
of the `red' LOSVD component coincides with the position of the knot K, within the
limits of 1 arcsec. In the following we will identify the `blue' component
as the main galaxy, and the second component as the `satellite'.

The analysis of the velocity fields of the main galaxy and the satellite was
carried out by means of the method of the `tilted-rings' model \citep{beg89}.
More details about this method, in connection to the MPFS velocity
fields, are given in \citet{moi04}.
The average parameters of the orientation of the rotated discs are
listed in Table~\ref{kinval}. The relatively high error in the system velocity
of the satellite is caused by an observed smooth change of
$V_{sys}$ with radius. Most likely, it is connected to the influence of
the tidal interaction on the dynamics of the satellite. Indeed, in the azimuthal
Fourier-decomposition of the velocity field there is a systematic
change of the harmonic with m = 1, which is interpreted in the frame of
the circular rotation as a radial trend of the systemic velocity.

\begin{table}
\caption{Kinematic parameters of the stellar components.}
\label{kinval}
\begin{tabular}{@{}llll@{}}
\hline
Name & $V_{sys}(Hel) $ & PA$_{dyn} $ & $i$  \\
& (\kms) & ($^{\circ}$) & ($^{\circ}$) \\
\hline main galaxy  & 11517 $\pm $ 9 & 57 $\pm$ 4 & 34 $\pm$ 3\\
satellite & 12164 $\pm$ 22 & 2 $\pm$ 6 & 35 $\pm$ 10   \\
\hline
\end{tabular}
\end{table}

In Fig.~\ref{starvel} the curves of the circular rotation for each of the
galaxies, and the azimuthally averaged radial distribution of the velocity
dispersion are also shown. We observe a rather small velocity dispersion of the
stars in Mrk 315, in comparison with a large amplitude of the rotation curve,
which does not reach a plateau within the limit of 5 kpc from the centre.
The ratio of the maximum velocity of rotation to the maximum velocity
dispersion, $V_{max}/\sigma \sim 3.5$, extremely exceeds the mean value which
is usually observed in the bulges of the early type galaxies
\citep[0.1 - 1.5,][]{korm82}. Moreover, there are no significant deviations of
line-of-sight velocity of the galaxy from the pure circular rotation.
Therefore, the stellar kinematics in Mrk 315 support the idea that we
are observing rotation of stars in a disc, instead of a dynamically `hot'
bulge.

In the centre a sharp decrease of the velocity dispersion of stars is
also observed, with values less than 50-70 \kms, which cannot be
measured with our spectral resolution. Recently \citet{woz03} explained such
drops in the radial velocity dispersion distribution of a number of Seyfert
galaxies, within the framework of the self-consistent dynamic model.
According to these authors, this effect is caused by stars which were born in
the centre from a dynamically cold gas having a smaller velocity dispersion,
than the older star population. Wozniak's  model has been constructed for
galaxies with bars, but in our opinion it can be  applied also to
Mrk 315, where the gas in the centre can be connected to tidal effects,
instead of a secular evolution of a bar.
Also, as mentioned in Section 7$a$, the bar could be existed here, but
dissolved by interaction.

Thus, our analysis of stellar kinematics in Mrk 315 clearly distinguishes two
independently rotating subsystems separated in systemic velocity by more than
$\sim 600$ \kms.
Estimating the masses of these stellar systems within the field-of-view by
means of the virial approximation $R \times V^2$, we obtain that the satellite
has a mass about 10 times lower than the mass of the main galaxy.
Therefore, our data demonstrate that the knot K is the debris of a nucleated
dwarf galaxy companion of Mrk 315, which experienced a minor merger episode.

The stellar velocity in knot K and the gas velocity in J1 and J2 are not in
fully agreement with stellar velocities of the satellite, but at
least within 150-200 \kms. They are not kinematically connected to the main
galaxy, but there could be a relationship between them. Since the mass of K is
largely smaller than the mass of the main galaxy, the collision should not
change significantly the morphology of the main galaxy, but it can compress
the gas and trigger the process of fuelling of the active nucleus.
Nevertheless, it is unlikely that such a small galaxy can throw away gas at the
observed velocity along J1.

\section[]{The environment}

To study the environment of Mrk 315, we applied the criteria
proposed by \citet{sch01} to consider a candidate companion as physically
bound to the active galaxy: 1) the distance to the main galaxy must be smaller
than 5 times the diameter of that galaxy; 2) the difference in brightness
between them must be $|\Delta m| \leq 3$ mag; 3) the difference
in radial velocities must be $|\Delta v| \leq 1000$ km s$^{-1}$.
On the DSS2-red image (Fig.~\ref{env}a) we found a galaxy located East of
Mrk 315 just at the distance of about 5 diameters ($\sim 3.4$ arcmin), which
satisfy also the brightness criterion ($|\Delta m| \sim 1.2$ mag).
This galaxy is catalogued in NED as 2MASX J23041747+2237302, but unfortunately
no information is available about its redshift.
Therefore, according to these criteria Mrk 315 can be considered a moderately
isolated galaxy.

Its overall morphology does not show strong distortions,
bright tidal tails or similar structures, which lead immediately to conclude
that this galaxy experienced past episodes of gravitational interactions.
Nevertheless, our analysis confirmed the presence of a secondary nucleus very
close to the active nucleus. This secondary nucleus originated likely by a
dwarf satellite which sank into the main galaxy.

Moreover, inspecting carefully the [O\,{\sc iii}] continuum-subtracted image,
we discovered an emitting source located $\sim 1$ arcmin South-East of
Mrk 315 (Fig.~\ref{env}b).
We measured the aperture magnitude V and R$_c$ of the active galaxy and
this source, obtaining V=14.60$\pm$0.09, R$_c$=13.92$\pm$0.07 (aperture diameter
= 100 arcsec), and V=20.0$\pm$0.3, R$_c$=19.7$\pm$0.2, respectively.
Clearly, the South-East source does not fit the brightness criterium.
Nevertheless, it is associated to a bright extended H\,{\sc i} cloud
\citep{sm01}.
This cloud could be an integral part of the outer regions of the main galaxy,
which is H\,{\sc i} deficient at velocities where the gas is ionised, that is
in the filament F1 or in the regions of starburst activity near the nucleus.
We obtained a low resolution slit spectrum of this source with SCORPIO
(Fig.~\ref{bcd}).
This spectrum shows prominent emission lines at a velocity
$\sim +200$ \kms relative to $V_{sys}$ of Mrk 315, and a clear stellar continuum.
Therefore this source, which has also a tidal-disturbed shape visible on the
broad-band images, is a dwarf satellite of the active galaxy.

We conclude that the criteria applied to evaluate the isolation of a galaxy are
often biased toward bright companions, and therefore are useless in case of
an environment populated by dwarf galaxies.
During the last two decades many authors have studied the environment of
active galaxies, mainly using a statistical approach and applying different
isolation criteria to define a companion galaxy. For many reasons, as for
example the crucial point of the control sample selection, they reached
contradictory results \citep[see e.g.][and references therein]{sch01}.
Therefore, to date a solid evidence that nuclear activity is induced by the
environment does not exist. On the contrary less attention has been paid to
dwarf satellites in relation to minor merger events.
Recently, on the basis of the Sload Digital Sky Survey (SDSS) EDR \citet{mil03}
observed a high fraction of AGNs ($\sim 40$ percent) in their sample of
nearby galaxies. They explained their result with either an AGN duty cycle 
longer than previous estimates by other authors, or with several bursts of the
nucleus driven by mergers.
Since they did not observe any dependence on environment of their AGN sample,
they rejected the second hypothesis.
Again, this analysis is limited to bright galaxies (M$(r^*)<-20$), and does not
take into account the possible presence of dwarf satellites. Indeed, converting
our magnitudes into the SDSS photometric system \citep{fuk96}, we obtained
M$(r^{\prime})=-21.8$ for Mrk 315 and M$(r^{\prime})=-16.1$ for the South-East 
companion.

\citet{derob98} and \citet{tan99} suggested that
`minor mergers' between a gas--rich galaxy and a satellite companion may
play a significant role in triggering activity in Seyfert nuclei.
Indeed the minor merger seems to be the favourite mechanism for several
reasons. First of all there is evidence that most spiral galaxies have dwarf
satellites \citep{zar97} and therefore minor merger events are
expected to occur several times during the lifetime of a galaxy.
Second, numerical simulations \citep[see e.g.][]{hm95} showed that
minor mergers can drive sufficient amount of gas from the host galaxy into
its central kiloparsec in a relatively rapid timescale ($<$ 1 Gyr).
Third, minor mergers do not cause strong deformations of the host morphology,
and indeed most Seyfert galaxies do not appear significantly different from
non--active galaxies. In fact, as reported by \citet{tan99}, the minor
merger timescale could be long enough to smear its relics, therefore most of
the advanced mergers would be observed as ordinary--looking isolated galaxies.

\section[]{Discussion}

Mrk 315 has been investigated by means of both new imaging and spectroscopic
data. The galaxy has several peculiar features in spite of the fact that it
appears like a rather `normal' early-type spiral. We discuss separately these
features in the following.\\

a) {\it Redshift}

We have compared the heliocentric systemic velocity of Mrk 315,
$V_{sys}=11517\pm9$ \kms, obtained with the spectroscopy of the stellar
component, with the values given in literature.
By inspecting carefully the previously published papers about this galaxy, we
found a redshift of 11827 \kms\ given by \cite{sar70} and \cite{hs73} based on
measurements of optical emission lines. We remark that this value is referred to
the Galactic centre, and when converted to heliocentric reference frame, it becomes
$\sim 11640$ \kms. The discrepancy with our redshift could be caused by the use
of the emission lines. In fact, the nuclear spectrum shows hydrogen Balmer lines
with asymmetric profiles caused by their broad components, which make
them useless to measure a systemic velocity, and high ionisation lines affected
by the nuclear outflow.
A value of 11827 \kms, coincidentally identical to the one given above, but in this
case referred to heliocentric reference frame, was obtained by \cite{mw84} with
H\,{\sc i} observations. The large difference with optical measurements could be
caused by the low spatial resolution of their data (3.3 arcmin). In fact,
this beam includes the South-East dwarf galaxy, which is an H\,{\sc i} source
much stronger than Mrk 315 (with a flux about nine times higher). This appears clear
in higher resolution H\,{\sc i} data recently published by \citet{sm01}, and
showing well separated H\,{\sc i} sources with velocities of $\sim 11800$ \kms\
and $\sim 11670$ \kms\ for the South-East dwarf and the main galaxy
respectively. Finally, the weak H\,{\sc i} cloud relative to Mrk 315 appears not
perfectly centred on its nucleus and extended toward North-West. This could
have been caused by gravitational interaction effects, and could justify the
discrepancy between the radio data and our systemic velocity measurement.

b) {\it Morphology}

At first sight the morphology of Mrk 315 does not show evident signs of past or
ongoing interaction.
Apart from the peak of ellipticity close to the nucleus and caused
by the presence of the knot, named K throughout this paper, the isophotes of
the galaxy are almost regular in their ellipticity. However, in all
analysed images after the subtraction of the 2-D surface brightness models,
two faint spiral arms are visible in external parts. The presence of these
spiral arms is likely to cause the observed change of the position angle
of the external isophotes.

The 2-D decomposition process of the surface brightness distribution allowed
us to give the correct interpretation of the morphological parameters observed
in the galaxy.
At the beginning we have considered two possible choices for the orientation
of the disc component in Mrk 315: 1) $PA_d \equiv PA_{out}\sim20^\circ$ and
$i_d \equiv i_{out}\sim30^\circ$, obtained from the parameters of the
external elliptical isophotes between 10 and 20 kpc;
2) $PA_d \equiv PA_{dyn}=57^\circ$ and $i_d \equiv i_{dyn}=34^\circ$, measured
on the velocity field of stars between 1 and 6 kpc;

In the first case (Fig.~\ref{mod} top) the residual brightness is negative
(over-subtracted) at $PA \sim 0^\circ$ and positive at
$PA\sim60^\circ-70^\circ$, where the
knot K is located, and resemble that of an elongated
bar-like feature in the galactic plane. But if this is a real bar, then the
non-circular motions of stars must shift the $PA_{dyn}$ in the gas/stellar
velocity field from the $PA$ of the line-of-nodes toward the opposite direction
relative to the bar major axis
\citep[see discussion and references in ][]{moi04}.
However $PA_{dyn}=57^\circ$, so it is turned to the same direction of the
bar major axis, like in case of disc kinematic.
Therefore the bar-like structure is an artifact.
Moreover, the morphological type of the galaxy is not consistent with the model
result. On one hand, the bulge-to-disc (B/D) luminosity ratio is
$L_B/L_D\sim2.5$ as in a E-S0 galaxy \citep[see ][]{sd86}.
On the other hand the bulge follows
an almost exponential profile (index $n\sim1.5$), which is rather strange for
an early-type galaxy \citep[see e.g. ][]{mgh98}.

On the contrary, the morphological structure becomes clear if we consider the
second choice for the disc orientation, that is by applying our kinematic
measurements. Firstly, the B/D ratio becomes $L_B/L_D\sim0.5-0.6$, which
corresponds to a Sab type according to \citet{sd86},
and the S\'ersic index $n\sim3$ is in a good agreement with this early
morphological type. Secondly, there is no more a displacement between the
stellar velocity field and the orientation of the disc. Thirdly, the map of the
residual brightness deprojected onto the galactic plane (see Fig.~\ref{mod}
bottom) shows a symmetric ring-like feature within 6 arcsec, in agreement with
the disc-like kinematics of this region, which remains even when an outer
secondary exponential disc is taken into account. This feature matches the
star-forming quasi-ring analysed in Section 4.
Finally, the presence of an inner disc allows us to explain the relatively
small value of the velocity dispersion of stars within a radius of 5 kpc (see
Section 5.3). Because this inner disc has a surface brightness greater or
equal to the bulge one (between 2 and 8 kpc), the dynamically cold component
influences significantly the line-of-sight velocity dispersion.

In this frame, the significant twist of the outer isophotes can be
explained as a warp of the outer disc. Substituting the parameters of the outer
isophotes $PA_{out}=20^\circ-30^\circ$ and $i_{out}=30^\circ$ in the equation (2)
by \citet{moi04}, we predict that the angle between the inner ($<8-10$ kpc) and
outer (15-20 kpc) parts of the galaxy is $\Delta i \sim 15^\circ-20^\circ$.
This result is consistent with a galaxy which experienced an interaction process
\citep[see e.g. ][]{rc99}, and therefore the presence of a warped disc is not
surprising.

In the last years a multicomponent structure of the galactic disc was found in
numerous galaxies, e.g. NGC 615 \citep{sva01} or the intriguing case of NGC 7217
\citep{sa00} with three exponential discs increasing
their radial scales outward. Also \citet{piz02} showed
nuclear discs embedded in the bulges of three early-type galaxies.
\citet{erw03} found luminous inner discs inside bar-regions of two barred
galaxies, NGC 2787 and NGC 3945, and in the last one the inner disc is ten times
more luminous than the bulge. \citet{moi04} also detected inner discs within a
one kiloparsec-size region of seven barred galaxies. To date the origin of
these inner discs is explained with a `secular evolution' of the
galactic disc, or a redistribution of the gaseous matter provoked by a
bar, or a moderate gravitational interaction with other galaxies
\citep[see discussion in ][]{sva01}.
Mrk 315 has no bar, but we found that it experienced interaction with two
dwarf satellites. In addition our model-subtracted images show a ring in
the central region (with the same scale of the inner disc) with two
bright peaks. The eastern peak is a satellite sank into the galaxy (knot
K), but the western peak belongs to the galactic plane, and therefore the
azimuthal brightness distribution of the ring is non-uniform. This
situation is analog to the results of numerical simulations of the
flight of a small satellite through a barred galaxy performed by
\citet{apb97}. Their Fig.~19 shows that the impact destroys the bar and form a
ring on a timescale $\sim 5 \times 10^7$ yr after the impact.
Therefore, we suggest that the inner disc component in Mrk 315 could be the
debris of a bar.

c) {\it Quasi-ring}

In analogy with \citet{non98}, who applied the method of isophotal fitting with
ellipses to make a model of the galaxy and subtract it from the original image
to evidence fine structures, we were able to enhance an internal quasi-ring
structure, embedded in the bright diffuse emission of the galaxy, and in
which we have identified three main regions (labelled A1, A2, and
A3). This structure, whose existence was already indicated by \citet{mac94} as
a ring, is similar to that shown in Fig. 1 of the \citet{non98}
paper, who called it a `chain' of knots surrounding the Seyfert
nucleus. We stress that we do not observe the `emission dip' found by these
authors at PA $\sim$ 225\degr, which should be located roughly in the middle
of A1.

This quasi-ring is characterised by diffuse H$\alpha$ emission, without
any evident condensation, while [O\,{\sc iii}] is very faint suggesting the
thermal origin of the ionising sources.
Indeed, integral-field spectra of regions A1, A2 and A3 exhibit the typical
emission lines and line ratios observed in H\,{\sc ii} and star
forming regions: low values of [N\,{\sc ii}]/H$\alpha$ ($<$ 0.6), [O\,{\sc
iii}]/H$\beta$ ($<$ 3) and high values of [O\,{\sc ii}]/[O\,{\sc iii}] ($>$
1).
This result is in agreement with \citet{wil88}, who measured
[O\,{\sc iii}]/H$\beta=0.4-1.0$ and claimed for hot star
photoionised gas outside the nucleus, where the kinematics is dominated by
circular motions.

The total H$\alpha$ luminosity, corrected for internal reddening, and then
converted into SFR clearly indicates that Mrk 315 is a starburst galaxy.
Our result is solid since we have cross-checked the high SFR value found by
using H$\alpha$ with infrared and radio emission, which gave a substantial
agreement. It is remarkable that \citet{non98} claimed for an overall SFR of
$\rm \sim 1~M_{\odot}~yr^{-1}$ on the basis of IRAS data taken from \citet{mbb91}.
We tried to investigate the reason for this large difference by repeating the
calculations made by these authors, and we found a possible oversight in their
FIR luminosity value, which is lower by a factor of $\sim 10$.
Indeed \citet{non98}
are not able to explain why a so quite low FIR luminosity, and therefore SFR,
can be related to a high value of SN rate given by the radio emission.
Moreover they used both 60 and 100 $\mu$m IRAS fluxes to obtain FIR emission,
but while at 60 $\mu$m the source is well defined and its emission easily
measurable, at 100 $\mu$m the galaxy is embedded in large and bright
structures which prevent a reliable estimate of the background level and
therefore of the source flux.
This is probably why the value at 100 $\mu$m  published by \citet{mbb91} is
an upper limit ($\leq 0.76$ Jy), and it is largely distant to the
flux given in the Faint Source Catalogue ($\leq 4.526$ Jy).

Through the analysis of the gaseous and stellar kinematics of
this structure, and more in general of the galaxy, we found out that
the velocity fields are made by independent kinematical components.
In particular, the main component of the galaxy is characterised by almost
pure circular rotation, with the dynamic centre located at the photometric
nucleus of Mrk 315.
This is a major difference between our results and those
presented by \citet{mac94}, and later mentioned by \citet{non98}, who located
the kinematic centre 6 arcsec NW of the nucleus.
In our opinion these authors might have been misleaded by the fact that their
long-slit spectrum was positioned close to kinematic minor axis, where rotation
is generally not expected, and covered partially the NW and SE region of high
velocity gas visible in the velocity field. This could be the reason of the
shape of the velocity curve obtained by \citet{mac94}.\\

d) {\it Knot}

A knot located close to the nucleus has been identified on the broad band
images, and named K throughout the paper.
It corresponds to the secondary nucleus of the merger
hypothesis claimed by MacKenty, who showed its stellar nature by means of an
I-band HST image.
The kinematical analysis of our integral-field data confirmed the merger
hypothesis: the knot K is a real secondary nucleus, the remnant of a dwarf
galaxy which sank into the main disc of the galaxy.
In fact, a stellar component kinematically independent from the main galaxy
was discovered and its
velocity field reconstructed by means of high resolution integral-field spectra
obtained with MPFS. The velocity field clearly shows a rotation pattern lying
over the knot K, which has a $\Delta V \sim +600$ \kms\ with respect to the
systemic velocity of the galaxy.

The high velocity of the impact is expected to produce gas compression.
Indeed, this secondary nucleus is embedded in the H$\alpha$ emission of
the galaxy, and it also shows  a faint [O\,{\sc iii}] emission, but is
strongly emphasised in the integral-field maps of [O\,{\sc iii}] $\lambda$4363,
[O\,{\sc i}] $\lambda$6300 and [S\,{\sc ii}] $\lambda$6724 emission lines.
All these features are a probe that the gaseous component at the location of
K is mainly ionised by shocks.
In particular, the auroral line [O\,{\sc iii}] $\lambda$4363 ($^1S_0-^1D_2$) can
have two main origins.
The collisionally populated level $^1S_0$ can be excited both by high
electron density, or by high gas temperature. Since the critical density of
this transition is around $\rm 10^8-10^9~cm^{-3}$, in case of lower density gas,
as it happens in NLR or generally in H\,{\sc ii} regions, this line is a
function of the temperature. In fact the
[O\,{\sc iii}] 4959+5007/4363 ratio is used as a direct measure of the gas
temperature, provided that the electron density $n_e < 10^6 ~\rm cm^{-3}$.
Therefore [O\,{\sc iii}] $\lambda$4363 is so weak in classical H\,{\sc ii}
regions, with temperatures ranging around 5000--10000 K, that is generally not
observed,  while it is visible when T $\sim 10^5-10^6$ K.
These are the conditions of supernova remnants, or more in general of shock
excited gas.

We stress that the knot K and its surroundings are characterised by higher
electron density and internal extinction than the other regions except
the nucleus. Moreover, the diagnostic line ratios indicate this secondary
nucleus shares the same properties observed in LINER galaxies.
In particular, it satisfies the [O\,{\sc ii}] $\lambda$3727/[O\,{\sc iii}]
$\lambda$5007 $\ge 1$, and [O\,{\sc i}] $\lambda$6300/[O\,{\sc iii}]
$\lambda$5007 $\ge 1/3$, emission line ratios criteria proposed by \citet{heck80}
and extensively discussed by \citet{ho04} to define a LINER.
Therefore, we cannot exclude that this knot could be a low luminosity active
nucleus, and if this is the case, Mrk 315 can be numbered as a new case of
two active nuclei in merging. Analogous results were obtained by
\citet{rif01} for ESO 202-G 23, and by \citet{kom03} for NGC 6240.\\

e) {\it Filaments}

The nature of the filaments mentioned in Section 3 is controversial.
F1 and part of F2 were already found out and discussed by \citet{mac86} and
\citet{mac94}, who considered them as a single filament with a `hook' shape.
Moreover, \citet{mac86} could see F1 only in narrow-band filters, and not in
continuum light.
At the beginning we also obtained a similar result, but later, thanks to deeper
images, we could observe that the situation is completely different: there
exist two independent filaments crossing likely in projection, and they are
both visible in continuum light.

The shape of F1 could indicate that it is a tidal tail, but some evidences are
against this hypothesis. First of all, the gas velocity is much higher than what
expected in case of a kinematical connection to the main body of the galaxy, as
it usually happens with the classical tidal tails \citep[see e.g.][and
references therein]{hib96}. Second, radio observations did not find any trace
of H\,{\sc i} emission in correspondence to the filament.
On the other hand, the jet hypothesis was already discussed and discarded by
\citet{mac86}. Essentially, the lack of radio emission beyond the boundaries of
the galaxy excludes that the filament could be a jet.
Indeed, the 6 and 20 cm observations by \citet{non98} showed only a diffuse
emission coincident with the extended regions of star formation found in the
quasi-ring, and two bright knots corresponding to the active nucleus N and
the secondary nucleus K.
Similar narrow and long structures may occur in case of relativistic jets from
active nuclei. The strong magnetic field causes a collimated radio emission
with non-thermal spectrum, which should be aligned with its optical counterpart,
as it is observed in radiogalaxies.
Even if the relativistic jet were oriented directly toward us, so we could not
observe synchrotron emission, it should become broader at the end due to the
bow-shock. This is not the case of F1.

We propose the following possible interpretation for both F1 and F2:
they are likely the debris of the interaction between Mrk 315 and two dwarf
companions. In particular, one sank into the main galaxy and gave rise to a
minor merger event (knot K), while the other passed closed to Mrk 315 in a
sort of fly-by (South-East dwarf).
It has been demonstrated by simulations and observations, that a satellite
companion crossing through the halo of the main galaxy undergoes a tidal
interaction, which strips away its outer stars, and forms extended tails with
low surface brightness which are usually difficult to detect \citep{forbes03}.
This idea is close to the hypothesis proposed by \citet{mac94}
and \citet{sm01}, that the filament F1 could be
`...{\it a short-lived wake left behind a bow-shock formed by the passage of
the galaxy's nucleus through the neutral gas}'.

In F1 the gas trails along the trajectory followed by the dwarf
galaxy, and its velocity increases along the filament, in agreement with the
long-slit observations by \citet{mac94}. Moreover, its distribution should be
curved near the massive core, as we see in the middle-band [O\,{\sc iii}]
images.
\citet{mac86} already observed that extrapolating the filament into the galaxy
failed to intersect the nucleus by $\sim 1$ arcsec.
If we take in account the length of the right side of the filament ($\sim 70$
kpc in projected distance), the falling time is about $1.2 \times 10^8$
years. Since this value is 3-5 times shorter than the period of rotation of
Mrk 315, we expect that the filament is not strongly disturbed by the
differential rotation.

The emission line ratios available for the part of the filament within the disc
of the galaxy (region J1) show an high ionisation degree, probed by the [O\,{\sc
iii}]/H$\beta$ and [O\,{\sc ii}]/[O\,{\sc iii}] values, which is similar to
those observed in the active nucleus. It is straightforward to claim that
nuclear radiation is ionising the gas along the filament.
This was also suggested by \citet{wil88}, who measured
[O\,{\sc iii}]/H$\beta$ ratios larger than 3 (and up to 17) in the North-West
region of the galaxy.

But energy budget calculations provide a flux of nuclear ionising photons lower
than that requested to sustain a so high ionisation at large distance from the
central engine. Therefore, another additional source must be invoked.
Since the gas is moving at high velocity along J1, we should reasonably expect
that strong shock excitation occurs (Mach number $>$10), and due to the low
density of the gas, the collisions should cause strong [O\,{\sc iii}] emission
(in analogy with [O\,{\sc iii}] structures observed in radio galaxies).

\section[]{Summary}

In this paper we have presented, analysed and discussed new data about the
Seyfert 1.5 galaxy Mrk 315, which allowed to improve the overall knowledge of
this peculiar object. Apparently isolated and undisturbed, Mrk 315 is known to
hide interesting features, who induced previous authors to classify it as a
merger in act.
The analysis of the light distribution in different wavelength ranges of the
optical domain, combined with the integral-field spectroscopic information,
showed that the galaxy has an early-type spiral morphology, with an inner and
an outer disc. The inner disc shows bright star forming regions arranged in a
quasi-ring shape around the nucleus, and it hosts a secondary nucleus close to
the AGN, centre of mass of a dwarf galaxy remnant, which sank into the main
body of the galaxy, producing clear effects of gas compression and shock
ionisation.
Highly ionised and collimated gas is observed to move radially at high velocity
with respect to the rotation of the galaxy, and in agreement with the velocity
of the secondary nucleus. This gas is connected to a giant filament.
We have discussed the nature of this filament, excluding the tidal tail
and/or jet hypotheses, and proposing the idea of debris of the dwarf galaxy,
which passed through the halo of the main galaxy loosing stars and
gas. We also identified a second and more extended filament, to which we
ascribed a similar origin. This filament is clearly connected to an emission
line dwarf galaxy, which has a redshift in agreement with that of Mrk 315, and
is associated to a bright  H\,{\sc i} cloud.

\section*{Acknowledgments}
We are grateful to the anymous referee for precious comments and suggestions
which improved the quality of the paper. \\
We are also grateful to Prof. S. Simkin and Prof. V. Reshetnikov for  
useful discussions. \\
A. Moiseev thanks Russian Science Support Foundation.\\
Based on observations carried out at the 6m telescope
of the Special Astrophysical Observatory of the Russian Academy of
Sciences, operated under the financial support of the Science Department
of Russia (registration number 01-43).\\
This research has made use of the NASA/IPAC Extragalactic Database (NED) which
is operated by the Jet Propulsion Laboratory, California Institute of
Technology, under contract with the National Aeronautics and Space
Administration.\\
This research was partially based on data from the ING Archive.\\

\label{lastpage}

\clearpage

\clearpage

\begin{figure*}
{\huge INSERT FIGURE\_3.JPG + FIGURE\_4.JPG}
\vskip 30pt
\caption{2-D decomposition of the surface brightness
distribution in the R$_c$ band. Top row shows the results of a single disc 
model with $PA=PA_{out}$. Here the left panel shows the residuals (smoothed) 
after the model subtraction from the original image in the sky plane, 
the middle panel is the residual after the deprojection onto the galactic 
plane. The scale is in magnitudes, the cross marks the photometric centre, and 
the solid line indicates the direction of the line of nodes. The right panel
presents the mean profile averaged in annular ellipses. The
solid line is the averaged model profile, and the dashed lines are the 
averaged profiles of the photometric components convolved with seeing. 
Bottom row is the same, but for two discs model with $PA=PA_{dyn}$.} 
\vskip 30pt
\label{mod}
\end{figure*}

\begin{figure*}
{\huge INSERT FIGURE\_5.JPG + FIGURE\_6.JPG}
\vskip 30pt
\caption{The quasi-ring structure. Contours are the isointensities of the 
non-calibrated H$\alpha$ ($a$) and [O\,{\sc iii}] ($b$) 
continuum-subtracted images of Mrk 315. 
The cross marks the position of the knot identified by \citet{mac94} 
as possible secondary nucleus.}
\vskip 30pt
\label{Ha+O3}
\end{figure*}

\begin{figure*}
{\huge INSERT FIGURE\_7.JPG}
\vskip 30pt
\caption{The V image of Mrk 315 with different greyscale levels to emphasise 
the weak inner (left panel) and outer (right panel) features.
In the left panel, black arrows indicate the position of the wounding trail, 
while the dashed line indicate the location of the inner stellar shell.
In the right panel, black arrows follow the wake toward the secondary nucleus 
(named F1 in the text), 
while white arrows indicate the wake toward the South-East extended source
(named F2 in the text).
The white dashed line indicates the position of the outer shell.
The scale--bar corresponds to 7.5 kpc at the distance of the galaxy.}
\vskip 30pt
\label{imaV}
\end{figure*}

\begin{figure*}
{\huge INSERT FIGURE\_8.JPG}
\vskip 30pt
\caption{Reconstructed maps of the galaxy for different emission lines in the 
field-of-view of the spectrograph (16$\times$15 arcsec). 
Each pixel corresponds to 1$\times$1 arcsec. The arrow indicates the 
North direction, the cross marks the position of the nucleus.}
\label{allmaps}
\end{figure*}

\clearpage

\begin{figure*}
\hbox{\includegraphics[width=8cm]{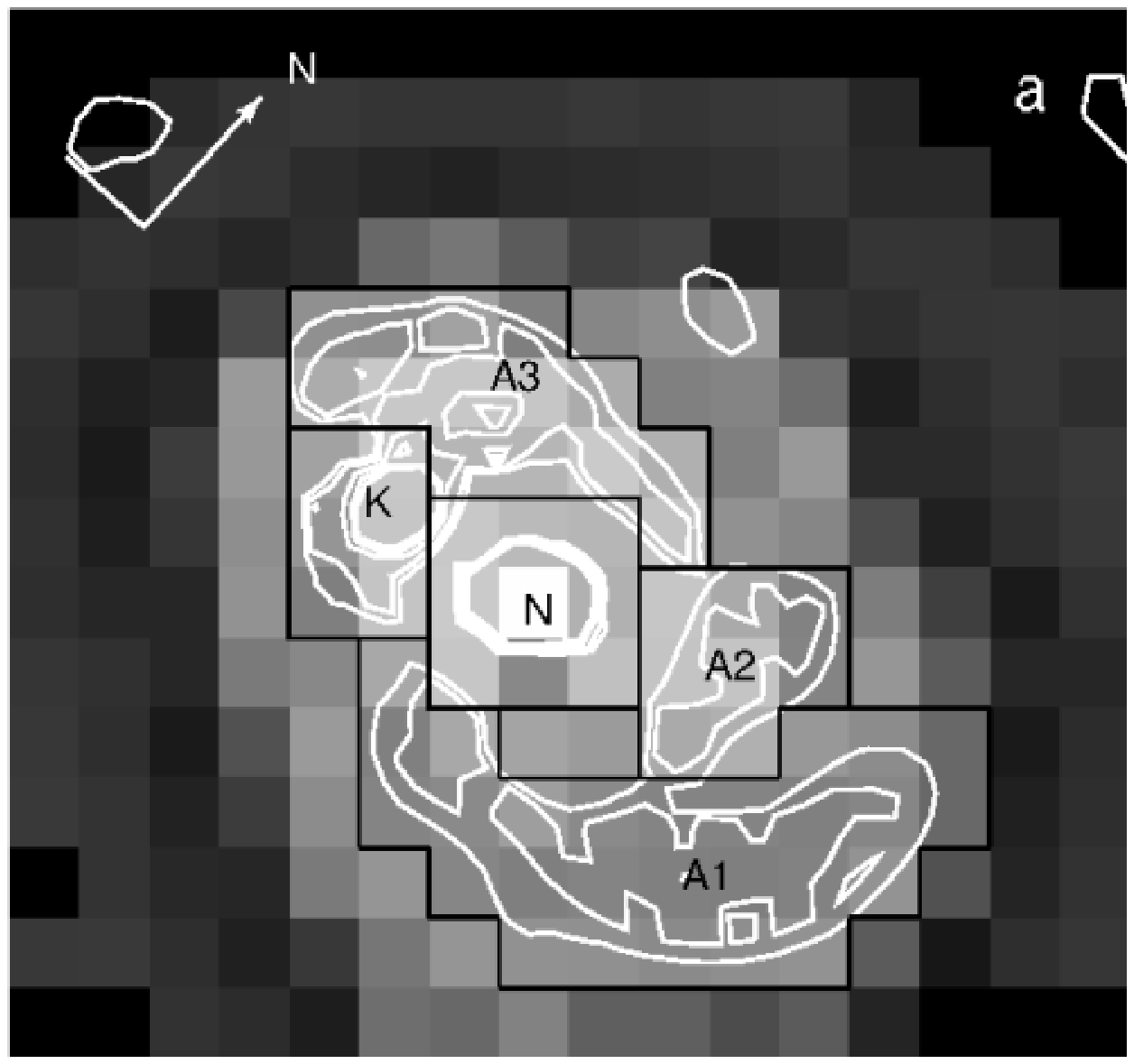}
\includegraphics[width=8cm]{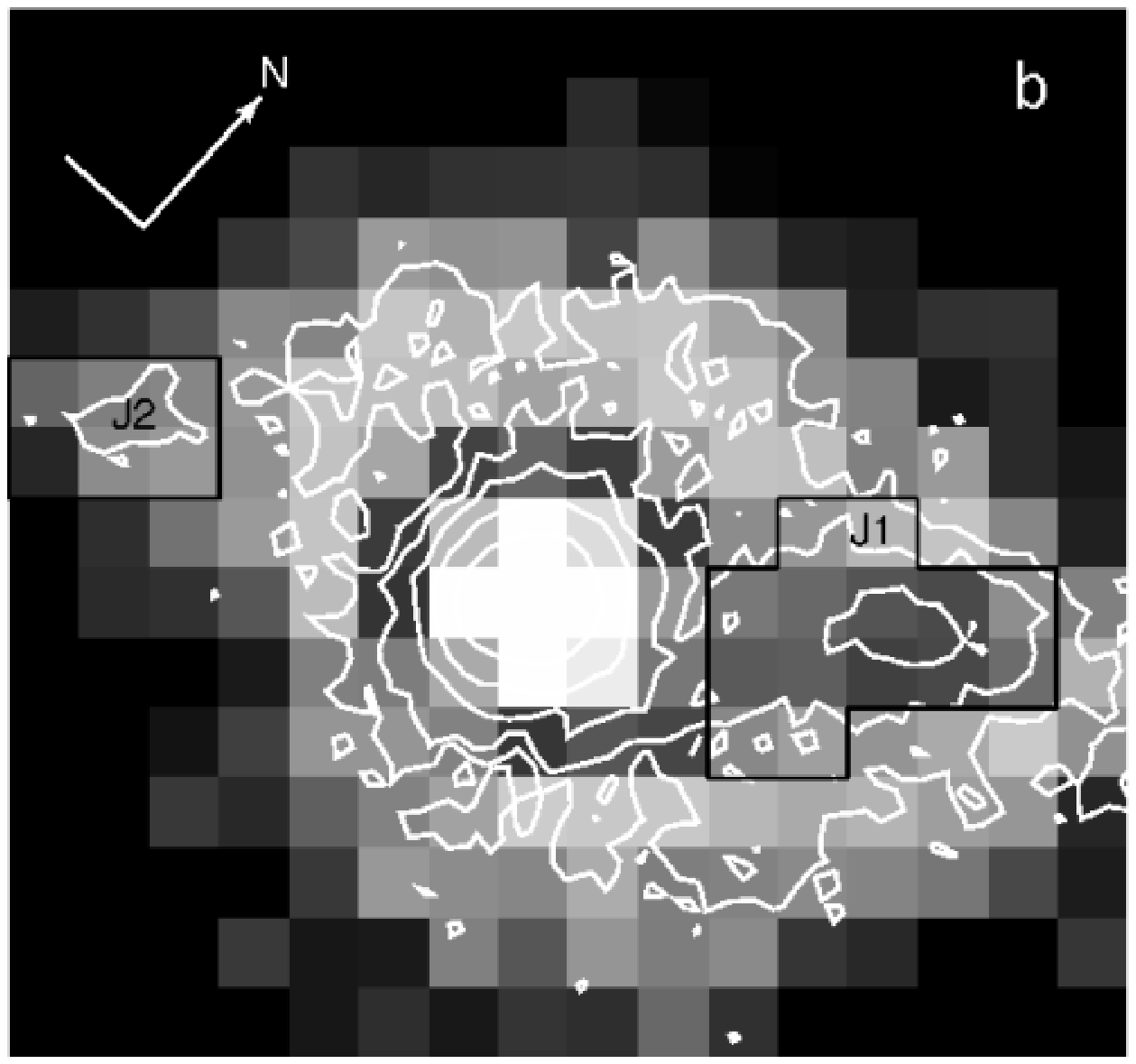}}
\caption{($a$) H$\alpha$ integral-field map. White contours are the 
isointensities of the V$_{res}$ image.
Black lines define the selected regions N, K, A1, A2 and A3.
($b$) [O\,{\sc iii}] integral-field map. White contours are the
isointensities of the [O\,{\sc iii}] continuum-subtracted image.
Black lines define the selected regions J1 and J2.
}
\label{regHa+O3}
\end{figure*}

\begin{figure*}
\includegraphics[width=13cm]{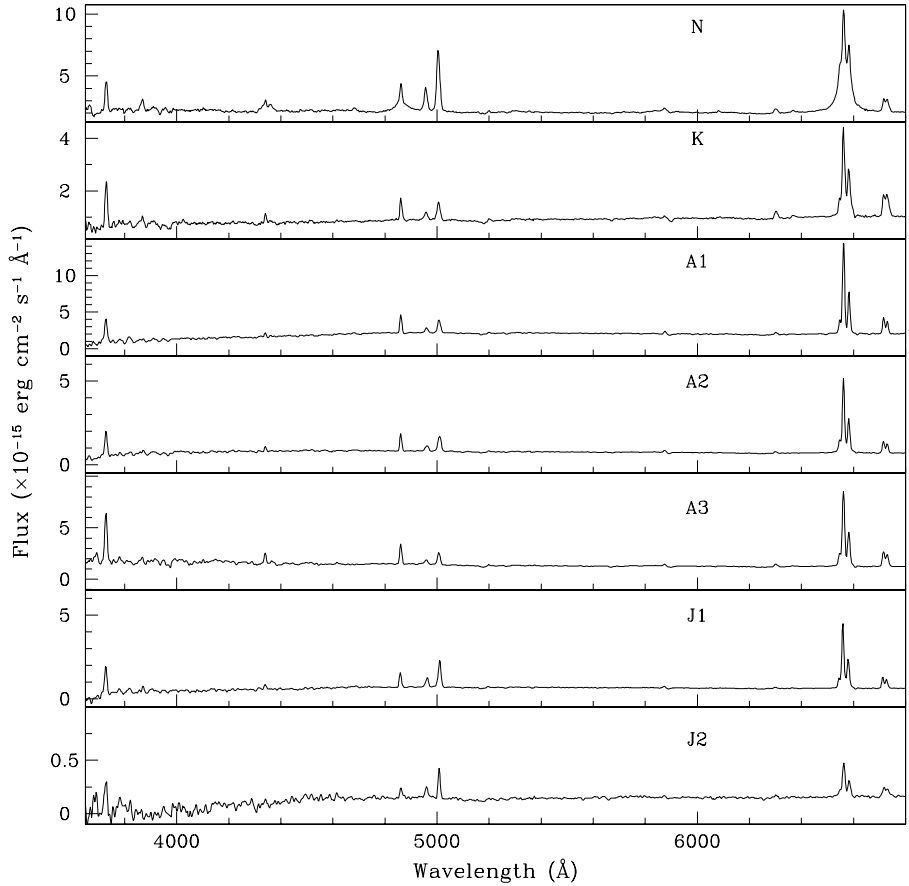}
\caption{Integrated spectra of the regions identified in Fig.~\ref{regHa+O3}.
}
\label{1d}
\end{figure*}

\begin{figure*}
{\huge INSERT FIGURE\_12.JPG}
\vskip 30pt
\caption{Integral-field maps of the emission line ratios used in diagnostic 
diagrams. The contours of the V$_{res}$ image are 
overlaid to all maps, 
but the [O\,{\sc iii}]/H$\beta$ where the contours of the 
[O\,{\sc iii}] continuum-subtracted image are used. 
The brightest pixels have ratios with the highest values.}
\vskip 30pt
\label{ratios}
\end{figure*}

\begin{figure*}
{\huge INSERT FIGURE\_13.JPG}
\vskip 30pt
\caption{
Map of the reddening corrected [O\,{\sc ii}]/[O\,{\sc iii}] ratio. 
The brightest pixels have the ratios with the highest values.
The contours of the [O\,{\sc iii}] continuum-subtracted image are overlaid.}
\vskip 30pt
\label{o2o3}
\end{figure*}

\clearpage

\begin{figure*}
\includegraphics[width=14cm]{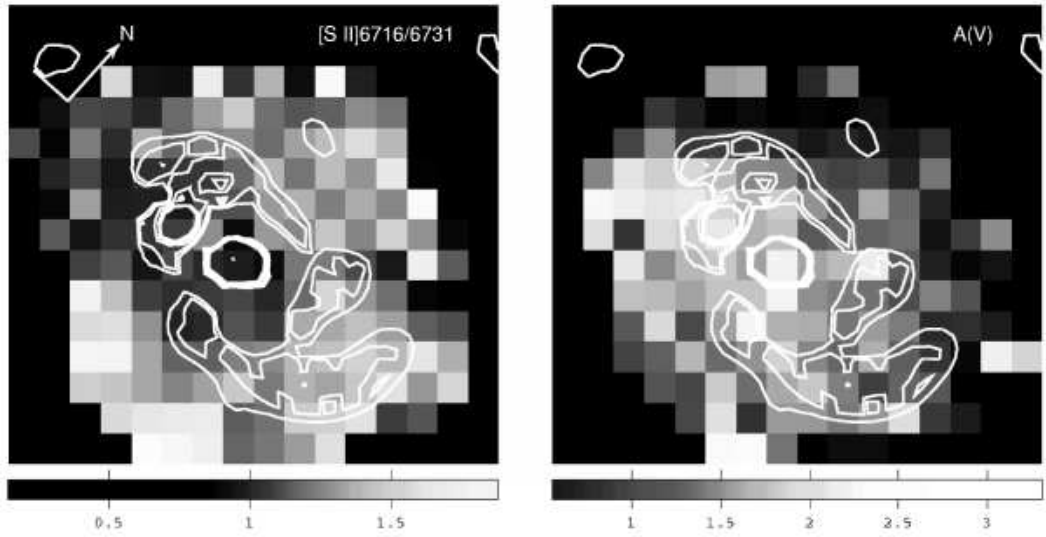}
\caption{{\it Left}:
Integral-field map of the [S\,{\sc ii}] doublet ratio, which indicates 
the distribution of the electron density ($n_e$) over the field of view.
Dark grey pixels correspond to the highest values of $n_e$.
{\it Right}:
The map of the internal reddening A(V) obtained from the H$\alpha$/H$\beta$ 
ratios. Brightest pixels are those with the highest extinction. 
White contours of the V$_{res}$ image are overlaid onto 
both panels.}
\label{de}
\end{figure*}

\begin{figure*}
\hbox{
\includegraphics[width=6.32cm]{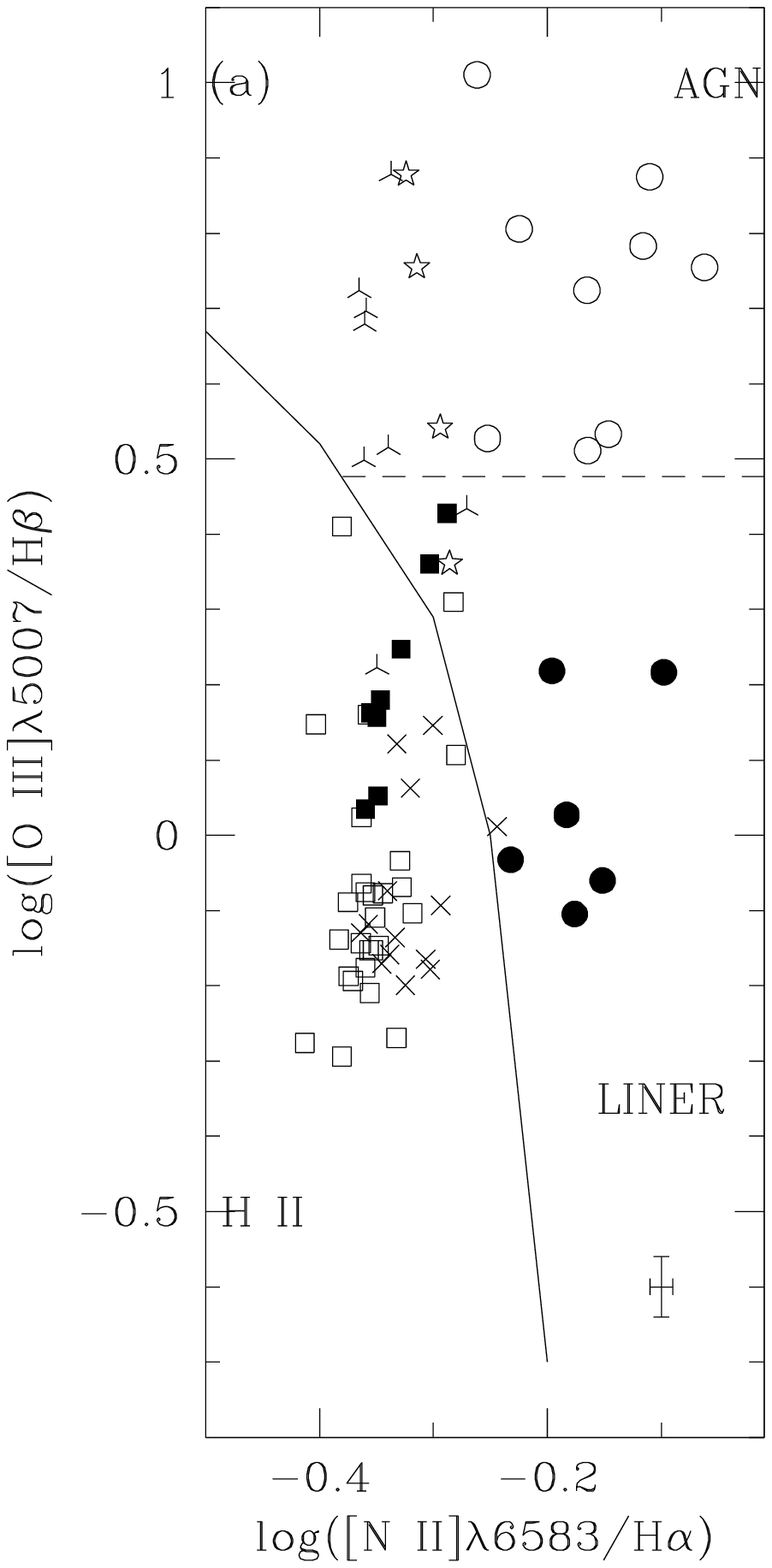}
\includegraphics[width=5cm]{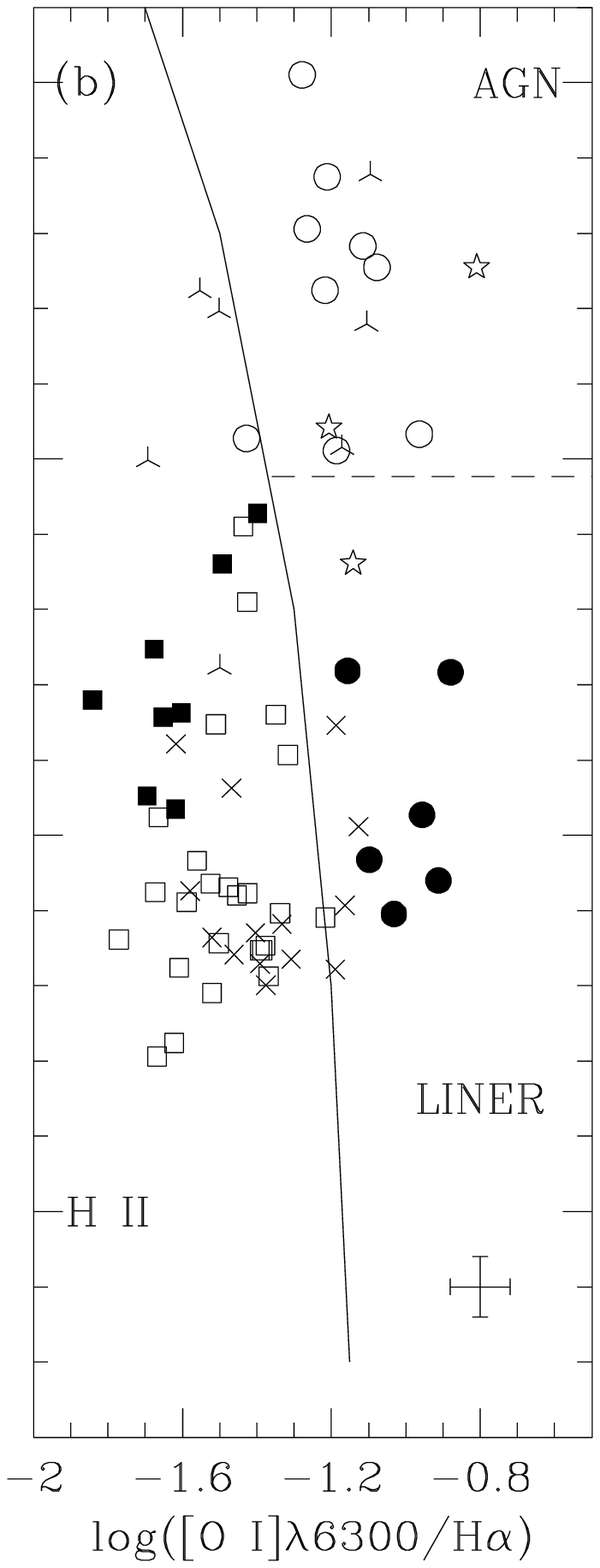}
\includegraphics[width=5cm]{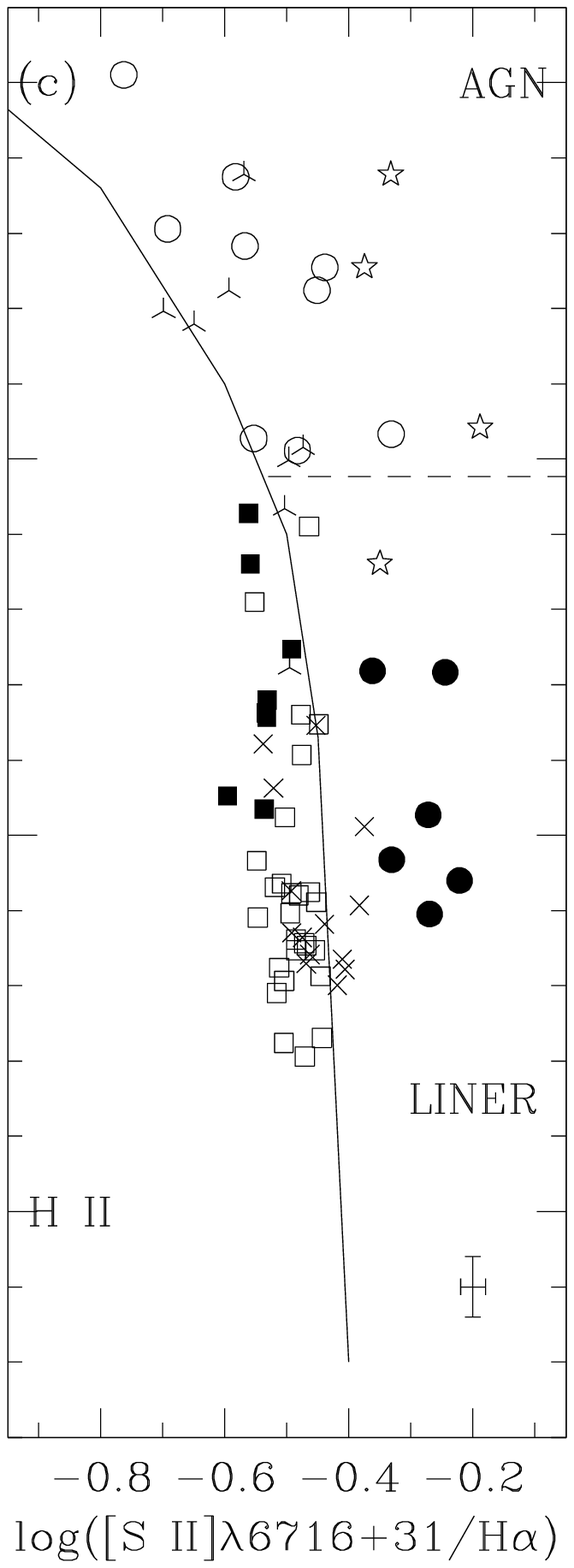}
}
\caption{\citet{vo87} diagnostic diagram showing the emission line ratios from 
the regions selected in Fig.~\ref{regHa+O3}: N (open circles), K (filled 
circles), A1 (open squares), A2 (filled squares), A3 (crosses), J1 (skeletal 
triangles), and J2 (open stars).
The solid and the dashed lines separate the zones where 
thermal, non-thermal and shock ionisation occurs.
The error bars indicate the typical errors for the diagnostic ratios of these 
regions.}
\label{vo}
\end{figure*}

\clearpage

\begin{figure*}
\includegraphics[width=15cm]{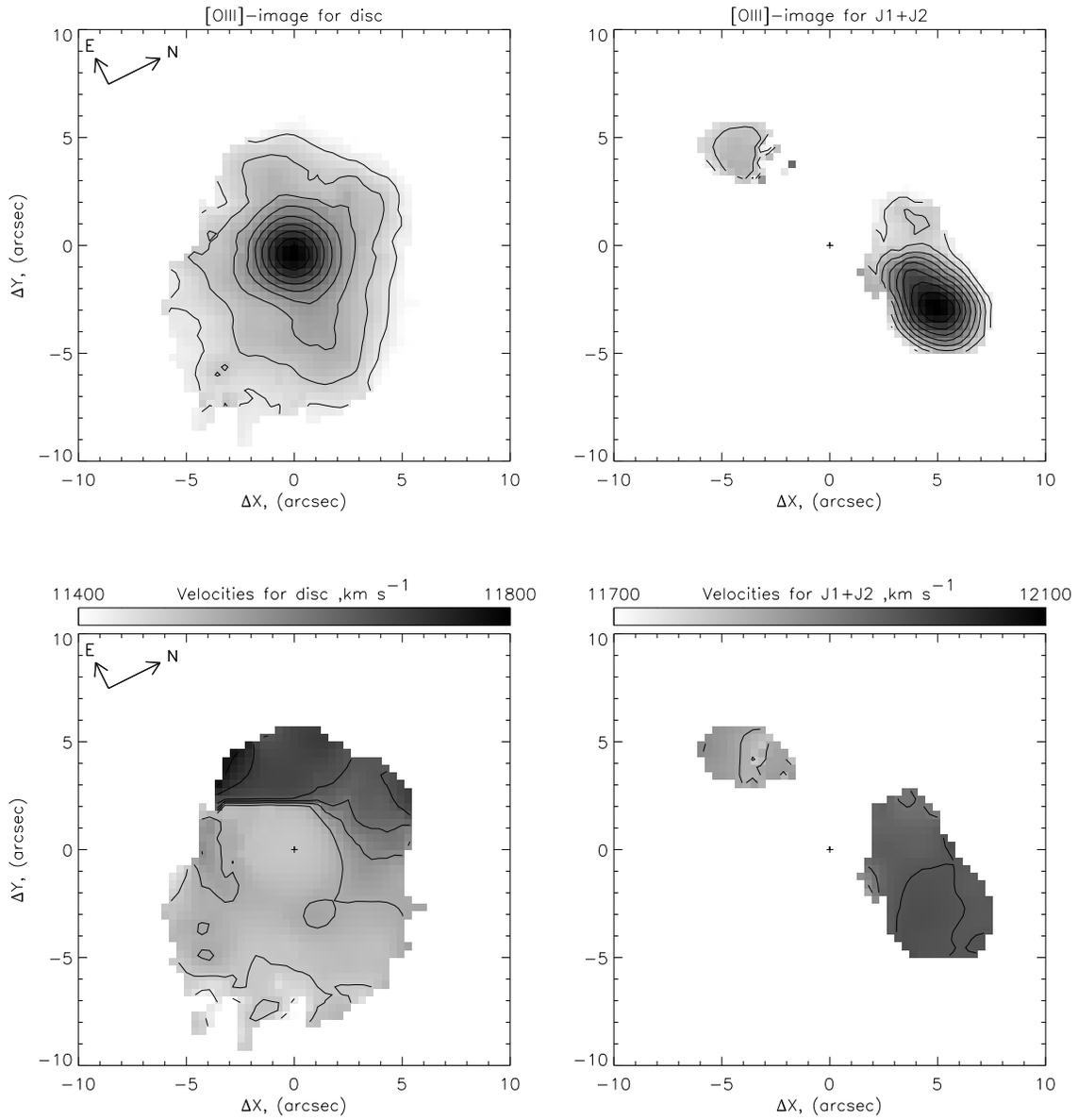}
\caption{The velocity field of gas obtained with Fabry-Perot observations 
in the emission line [O\,{\sc iii}]. Two kinematic components can be separately
studied. The left panel shows the rotation of the galaxy with the additional
nuclear outflow. The right panel shows the high velocity gas corresponding 
to the J1 and J2 regions.}
\label{o3vel_fp}
\end{figure*}

\begin{figure*}
\centerline{\includegraphics[width=15 cm]{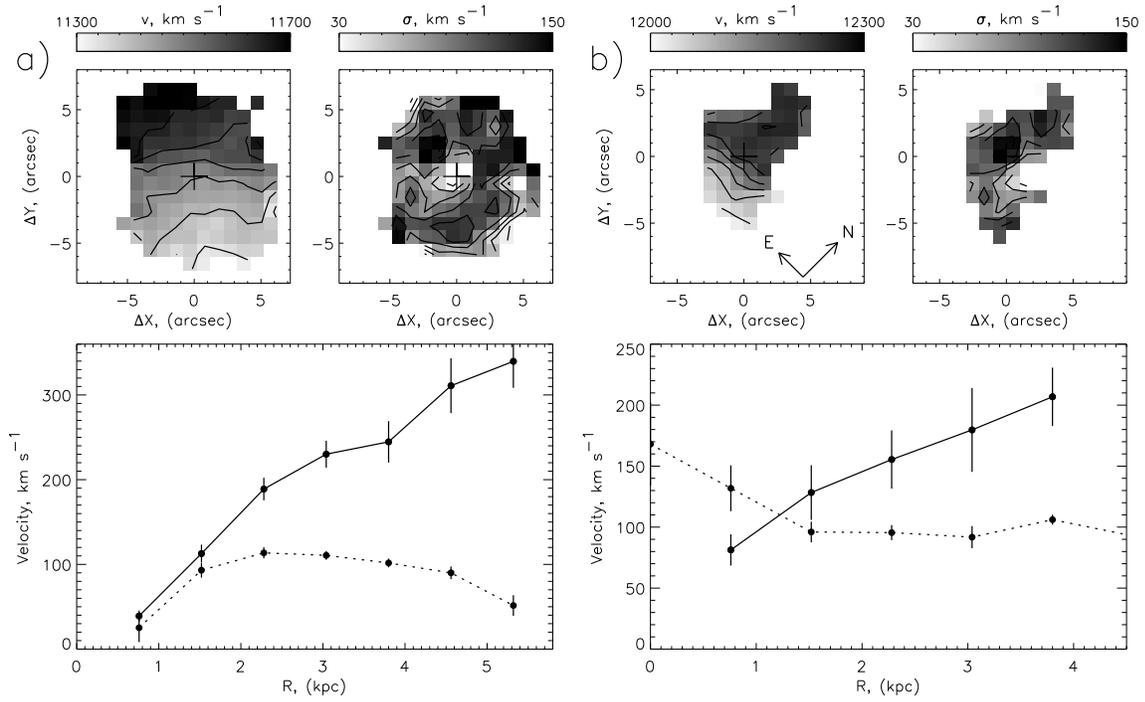}}
\caption{Stellar kinematics obtained with MPFS data. 
({\it a}) The `main galaxy' component: the line-of-sight
velocity field and the map of the velocity dispersion are reproduced in the
upper panels. The cross marks the dynamical centre. The rotation curve 
(solid line) and the average radial distribution of the
line-of-sight velocity dispersion (dotted line) are plotted in the lower panel.
({\it b}) The same for the secondary component, named `knot K' in the text.} 
\vskip 30pt
\label{starvel}
\end{figure*}

\begin{figure*}
{\huge INSERT FIGURE\_20.JPG}
\vskip 30pt
\caption{The POSS2-red field around Mrk 315 (panel (a)). The black and white 
arrows indicate the positions of Mrk 315 and the nearest brightest galaxy 
2MASX J23041747+2237302.
The [O\,{\sc iii}] continuum-subtracted image of Mrk 315 (black arrow) and 
the location of the dwarf emission-line galaxy (white arrow).}
\label{env}
\end{figure*}

\clearpage

\begin{figure*}
\includegraphics[width=15cm]{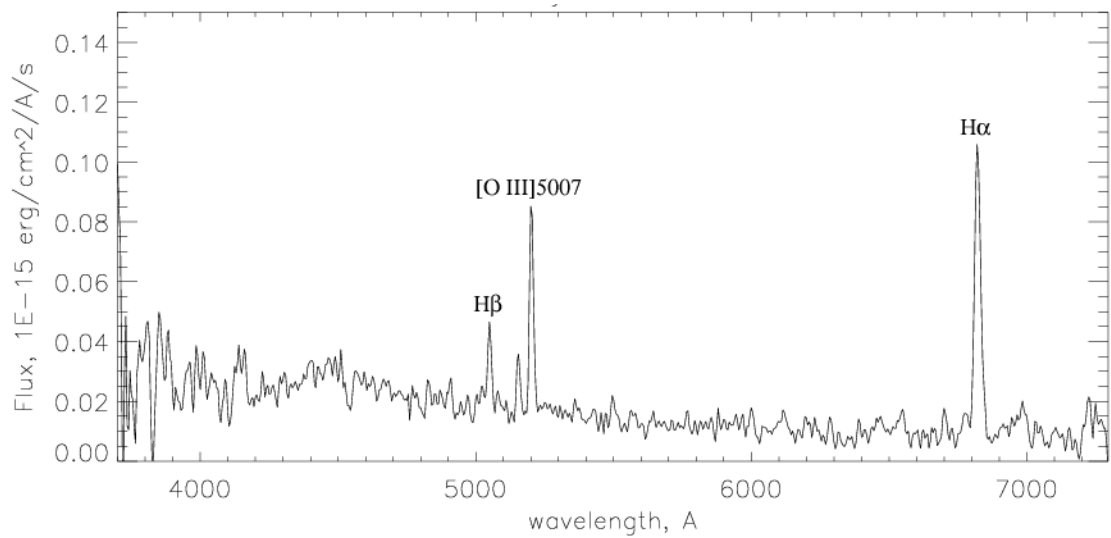}
\caption{The low resolution spectrum of the dwarf emission-line galaxy 
identified in [O\,{\sc iii}] with SCORPIO.}
\label{bcd}
\end{figure*}

\end{document}